\documentclass[aps,prc,twocolumn,groupedaddress,showpacs]{revtex4}

\usepackage{graphicx}% Include figure files
\usepackage{bm}% bold math

\begin{document}

\title{Canonical-basis solution of the Hartree-Fock-Bogoliubov equation
on three-dimensional Cartesian mesh}

\author{Naoki Tajima}
\email[]{tajima@quantum.apphy.fukui-u.ac.jp}
\homepage[]{http://serv.apphy.fukui-u.ac.jp/~tajima}
%\thanks{}
\affiliation{Department of Applied Physics, Fukui University}

\date{\today}

\begin{abstract}
  A method is presented to obtain the canonical-form solutions of the HFB
equation for atomic nuclei with zero-range interactions like the Skyrme
force.  
  It is appropriate to describe pairing correlations in the continuum
in coordinate-space representations.  
  An improved gradient method is used for faster convergences under
constraint of orthogonality between orbitals.  
  To prevent high-lying orbitals to shrink into a spatial point, a
repulsive momentum dependent force is introduced, 
  which turns out to unveil the nature of high-lying
canonical-basis orbitals.
  The asymptotic properties at large radius and the relation with
quasiparticle states are discussed for the obtained canonical basis.
\end{abstract}

%\keywords{}

\maketitle

%----------------------------------------------------------------------------

% Private commands which take no arguments
\newcommand{\etal}{{\em et al.\ }}
\newcommand{\vct}[1]{\bm{#1}}
\newcommand{\dvol}{d^3 r} % volume element
\newcommand{\fermilevel}{\epsilon_{\rm F}}
\newcommand{\porn}{{\rm q}} % label for proton or neutron
\newcommand{\itss}{\, \Delta t \,} % imaginary time step size
\newcommand{\ekinmax}{T_{\rm max}}  % {E_{\rm kin}^{\rm (max)}}
\newcommand{\nmesh}{N_{\rm mesh}}  % The number of mesh points in one direction
\newcommand{\eqpcut}{E_{\rm c}}  % Cut-off energy of quasiparticles
\newcommand{\nosc}{N_{\rm osc}}  % Principal quantum number of H.O. potential 
\newcommand{\alphaip}{\mu}
\newcommand{\betaip}{\nu}

% Figure numbers: seem necessary when they are referred to in figure captions.
\newcommand{\figcanorb}{3} % \ref{fig:canorb}
\newcommand{\figwftail}{4} % \ref{fig:wf_tail}
\newcommand{\fignwetot}{9} % \ref{fig:nw_etot}
\newcommand{\figshrink}{15} % \ref{fig:shrink}
\newcommand{\figshrinkkc}{16} % \ref{fig:shrink_kc}

%==============================================================================
\section{Introduction \label{sec:intro}}

\def\baselineskipTaj{\baselineskip}

% << Pairing in the continuum is necessary for half of nuclides. >>

Pairing correlations play an essential role in the determination of the
ground-state structure of the vast majority of atomic nuclei.  
Among its treatments, an easy but still enough accurate one is to
consider only single-particle states near the Fermi level while the
effects of the other states are assumed to be absorbed in the strength
of an effective pairing interaction.
In such a treatment,
our experiences in Hartree-Fock(HF)+BCS calculations suggest that
at least half of a major shell above the
Fermi level must be considered for meaningful estimations.
As a consequence, one has to explicitly consider positive-energy states
if the Fermi level of neutrons is higher than the negative of half of
the major shell spacing ($\sim \frac{41}{2} A^{-1/3}$ MeV).
This condition applies to about half of the $10^4$ nuclides 
which exist between proton and neutron drip lines in the nuclear chart,
not only to nuclei near the neutron drip line or outside the s-process path.
Thus there are plenty of necessities for
theoretical frameworks which can describe pairing correlations
involving the continuum states.

% << HF+BCS is not sufficient. HFB is necessary. >>

If the pairing correlation significantly involves
the continuum states, the HF+BCS
approximation becomes inadequate because the occupation of unbound
HF orbitals leads to the unwanted dislocalization of the nucleon density.
This is a serious problem in coordinate-space treatments, which is
more favorable to describe loosely bound systems like drip-line
nuclei than expansions in harmonic-oscillator eigenstates (except
the transformed oscillator basis of Ref.~\cite{SNP98}).
The reason why a dislocalized solution becomes the variational minimum is
that, in order to separate the variational equations into HF and BCS,
one has to neglect the effects that the matrix elements of
pair-scattering processes are affected by the changes in the 
wave functions of the orbitals involved \cite{VB72}.
To fully take into account these effects leads to the
Hartree-Fock-Bogoliubov (HFB) equation, with which
the density is localized whenever the Fermi levels are negative
\cite{DFT84,DNW96}.

The HFB method in the coordinate space was first formulated using
the quasiparticle states and solved for spherically symmetric states
in Ref.~\cite{DFT84}.
Although spherical solutions can be obtained easily with present
computers (for zero-range forces), deformed solutions are still
difficult to obtain because there are quite a large number of
quasiparticle states even for a moderate size of the normalization box
(i.e., the cavity to confine the nucleons to discretize the
positive-energy single-particle states).
An orthodox approach to face this difficulty is the two-basis method
\cite{GBD94,THF96,TFH97,Taj98a,YMM00}, in which the quasiparticle states are
expanded in bound and unbound HF orbitals. 
This method requires heavy numerical calculations because there are a pile of
unbound HF orbitals below an energy cut-off of even only a few MeV.

An alternative approach is the canonical-basis HFB method.  According to
the Bloch-Messiah theorem \cite{BM62}, every HFB solution obtained as the
vacuum of a set of Bogoliubov quasiparticles has an equivalent
expression in terms of a BCS-type wave function.  
The single-particle states appearing in this
expression are called the HFB canonical basis.  
In the canonical-basis HFB method, one obtains the solution in the
canonical form without using the quasiparticle states.
This method appeared originally in Ref.~\cite{RBR97} to obtain
spherical solutions. However, there are no sever difficulties in 
obtaining HFB solutions for spherical nuclei using any other methods.
Applications to deformed nuclei have been done by us
\cite{Taj98a,Taj98b,Taj99a,Taj00a}
using a three-dimensional Cartesian mesh representation\cite{BFH85}.
Incidentally, a different line of application is also found 
in literature \cite{ST01}.

\begin{figure}%G:psf
\includegraphics[angle=-90, width=8.6 cm]{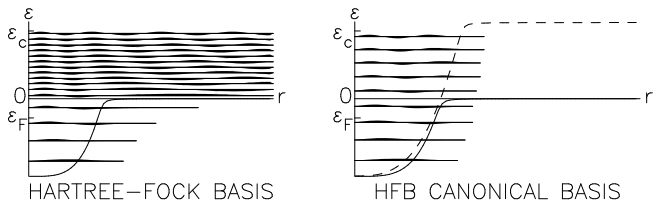}%G:psf
\caption{
Comparison between the HF orbitals and the HFB canonical-basis orbitals.
See text for explanations.
}
\label{fig:two_basis_HFB}
\end{figure}

Let us explain the advantage of the canonical-basis method 
over the two-basis method concerning the treatment
of the pairing in the continuum using Fig.~\ref{fig:two_basis_HFB}.
On both sides of the figure, the ordinate
represents the expectation value of the mean-field (HF) energy, while
the abscissa stands for the radius $r$ from the center of the nucleus.
$\epsilon_{\rm c}$ and $\epsilon_{\rm F}$ mean a cut-off energy and the Fermi
level, respectively.  On the left-hand side, the wavy horizontal lines
stand for energy levels and their spatial extent for the HF potential
denoted by a solid curve.  The wave functions of positive energy states
extend to the wall of the box and their level density is much larger
than that of negative energy states.  
In the two-basis HFB method, one has to mix these positive-energy
orbitals to construct localized canonical-basis orbitals, which is a
numerically demanding task.  On the right-hand side, the wavy lines
represent the HFB canonical-basis orbitals.  Unlike HF orbitals, they are
spatially localized for both negative and positive energies. Because of
this localization, the level density is much smaller than the unbound HF
orbitals.
Therefore, one needs much less orbitals.
More specifically,
the number of necessary single-particle states to describe the
HFB ground state of a nucleus is proportional to the volume of the nucleus
in the canonical-basis method while it is related to the volume
of the normalization box in the other methods.  
Incidentally, with the dash curve, we suggest the existence
of some potential which binds the high-lying canonical-basis orbitals. 
The identity of this potential is unveiled in Sec.~\ref{sec:canorb}.

%-------------------------- F I G U R E -----------------------------------
%\begin{figure}[h]%G:box
%\begin{center}\framebox{Figure \ref{fig:two_basis_HFB}}\end{center}%G:box
%------------------- E N D   O F   F I G U R E ----------------------------

In this paper we formulate the canonical-basis HFB method on a cubic
mesh and develop an efficient gradient method to obtain its solutions.
In order to decrease unphysical influences of high-momentum components
due to zero-range interactions, we introduce a momentum dependent term
to the pairing interaction and show how it suppresses a problematic
behavior of wave functions peculiar to the canonical-basis HFB method.
As a byproduct, we show a clear way to understand the nature of high-lying
canonical-basis orbitals.

The contents of this paper is as follows.
In sections \ref{sec:cbhfb} -- \ref{sec:gradientmethod}, 
we give the formulation and general considerations.
Discussions using results of numerical calculations
are collected in section \ref{sec:results}. 
Conclusions are given in section \ref{conclusions}.
In Appendix \ref{sec:meshprec} we examine the errors
due to the mesh representation and the box boundary condition.
Appendix \ref{sec:pointcollapse} gives a model analysis of the problem
peculiar to the method, i.e., a phenomenon that 
canonical-basis orbitals collapse to a spatial point
once their occupation probabilities fall below some critical value
when pure delta-function forces are used.

Some of the contents of this paper have already been discussed in our earlier 
reports \cite{Taj98a,Taj98b,Taj99a,Taj00a}.
For example, Ref.~\cite{Taj98a} includes the first report on
the canonical-basis HFB method on a cubic mesh.
Refs.~\cite{Taj99a,Taj00a} reported 
on the discovery of the point-collapse problem and
the tests of the state-dependent cut-off factors and
pairing-density-dependent forces as the remedies.
In this paper we have rewritten those contents
to include the momentum-dependent interaction and,
by including the rewritten contents,
organized the formalism for consistency and selfcontainedness.
All the numerical calculations shown in this paper have been performed 
using the renewed form of the interaction.

Incidentally, some important parts of those reports are not repeated in
this paper.  The examples are an analysis of the spherical two-basis
method concerning the precision of density tails \cite{Taj98a} and
numerical demonstrations of the point collapses of high-lying orbitals
\cite{Taj00a}.

%==============================================================================
\section{HFB in the canonical representation \label{sec:cbhfb}}

To begin with, let us formulate HF and HFB methods in the
coordinate-space representation in order to elucidate the difficulty of
the quasiparticle-based HFB method compared with the HF method and to
propose its possible solution in terms of the canonical-basis HFB method.
For the sake of simplicity, we consider only one kind of nucleons in
this section.  The number of that kind of nucleons are designated by
$N$.  The $z$-component of the spin of a nucleon is represented by $s$
(= $\pm \frac{1}{2}$).

In HF, one should minimize $\langle \Psi | H | \Psi \rangle$
among all the $N$-body single Slater-determinant states,
\begin{eqnarray}
\label{eq:vac_hf} % -*- Eq : vacuum of HF
|\Psi \rangle & = & \prod_{i=1}^{N} a^{\dagger}_i | 0 \rangle, \\
\label{eq:cre_op_hf} % -*- Eq : creation operator for HF s.p. orbital
a^{\dagger}_i & = & \sum_{s} \int \dvol \; \psi_{i} (\vct{r},s) \;
a^{\dagger}(\vct{r},s),
\end{eqnarray}
by varying $\{ \psi_i \} _{i=1,\cdots,N}$ under orthonormality
conditions $\langle \psi_i | \psi_j \rangle$ = $\delta_{ij}$.
The operator $a^{\dagger}_i$
creates a nucleon with a wave function $\psi_{i} (\vct{r},s)$.
The ket state $|0\rangle$ stands for the state in which no nucleons exist.
The distribution function of the number density of the nucleons
is related to the single-particle wave functions as
\begin{equation}
\label{eq:dens_hf} % -*- Eq : nucleon density operator and density for HF
\rho ( \vct{r} ) 
% = \sum_{s} \langle \Psi | a(\vct{r},s) a^{\dagger}(\vct{r},s) | \Psi \rangle
= \sum_{i=1}^{N} \sum_{s} | \psi_i ( \vct{r},s ) |^2.
\end{equation}
There is arbitrariness in the selection of $\{ \psi_i \}$ 
as for unitary transformations among them.
$\psi_i$ are called HF orbitals when they diagonalize the
HF Hamiltonian.

In HFB, the solution takes the following form,
\begin{eqnarray}
\label{eq:vac_hfbq} % -*- Eq : vacuum of HFB-Q (quasiparticle formalism)
|\Psi \rangle & = & \prod_{i=1}^{N_{\rm B}} b_i | 0 \rangle, \\
b_i & = & \sum_{s} \int \dvol \left\{
\phi_i^{\ast} (\vct{r},s) \; a(\vct{r},s)
+ \varphi_i^{\ast} (\vct{r},s) \; a^{\dagger}(\vct{r},s) \right\},
\label{eq:ani_op_qp} % -*- Eq : quasiparticle annihilation operator
\end{eqnarray}
where $b_i$ is the annihilation operator of a positive-energy
Bogoliubov quasiparticle with two types of amplitudes
$\phi_i (\vct{r},s )$ and $\varphi_i ( \vct{r}, s )$ 
for particle-like and hole-like parts of the excitation.
$N_{\rm B}$ is the number of the basis states of the representation used
and is by far larger than $N$. One should vary
$\{ \phi_i, \varphi_i \}_{i=1,\cdots,N_{\rm B}}$
under an orthonormality condition
\begin{equation}
\label{eq:norm_qp} % -*- Eq : normalization of quasiparticle w.f.
\sum_{s}\int
\left\{
    \phi_i^{\ast}   ( \vct{r}, s ) \phi_j    ( \vct{r}, s )
  + \varphi_i^{\ast}( \vct{r}, s ) \varphi_j ( \vct{r}, s )
\right\} \dvol
 = \delta_{ij} ,
\end{equation}
and a constraint on the expectation value of the number of nucleons,
\begin{equation}
\label{eq:num_hfbq} % -*- Eq : nucleon number of HFB-Q
\int \rho (\vct{r}) \dvol = N,
\end{equation}
where the nucleon density for state (\ref{eq:vac_hfbq}) is expressed as
\begin{equation}
\label{eq:dens_hfbq} % -*- Eq : nucleon density of HFB-Q
\rho ( \vct{r} ) = \sum_{i=1}^{N_{\rm B}} \sum_{s}
| \varphi_i ( \vct{r},s ) |^2.
\end{equation}

The essential difference between HF and HFB lies
in the number of necessary single-particle wave functions.
It is $N$ in the former and $2 N_{\rm B}$ in the latter.
Obviously, the latter is much larger.

Owing to the Bloch-Messiah theorem \cite{BM62},
the state (\ref{eq:vac_hfbq}) can be transformed
(as for the ordinary ground states of even-even nuclei) 
into the following form,
\begin{eqnarray}
\label{eq:vac_cbhfb} % -*- Eq : vacuum of canonical-orbital HFB
|\Psi \rangle & = & \prod_{i=1}^{N_{\rm w}}
\left( u_i + v_i \; a^{\dagger}_{i} \; a^{\dagger}_{\bar{\imath}} \right)
| 0 \rangle,\\
\label{eq:cre_op_cbhfb} % -*- Eq : canonical-orbital creation operator
a^{\dagger}_{i(\bar{\imath})} & = & \sum_{s} \int \dvol \; 
\psi_{i(\bar{\imath})} (\vct{r},s) \;
a^{\dagger}(\vct{r},s),
\end{eqnarray}
where $a^{\dagger}_{i}$ and $a^{\dagger}_{\bar{\imath}}$ create a
nucleon with wave functions $\psi_i(\vct{r},s)$ and
$\psi_{\bar{\imath}}(\vct{r},s)$, respectively, which are called
the canonical basis of the HFB vacuum $| \Psi \rangle$.
They may be called the natural orbitals\cite{RBR97} instead,
because they are the eigenstates of the one-body density matrix.
In this paper, we often call them canonical-basis orbitals or 
canonical orbitals.
It is required 
$N_{\rm w}=\frac{1}{2}N_{\rm B}$ for the exact equivalence
between Eqs.~(\ref{eq:vac_hfbq}) and (\ref{eq:vac_cbhfb}) to hold in general.

In this paper we assume that $| \Psi \rangle$ is 
invariant under the time reversal operation.
In this case, $\psi_i$ and $\psi_{\bar{\imath}}$ are the time-reversal 
state to each other and only one of them has to be considered explicitly
in the variational procedure.
The nucleon density can be expressed as
\begin{equation}
\label{eq:dens_cbhfb} % -*- Eq : nucleon density of HFB-N
\rho ( \vct{r} ) = 2 \sum_{i=1}^{N_{\rm w}} \sum_{s} v_i^2
\left| \psi_i ( \vct{r},s ) \right|^2,
\end{equation}
in the time-reversal invariant case.

In order to obtain solutions expressed in the canonical form
without using the information of the quasiparticle states, 
one should vary
$\{ \psi_i , u_i , v_i \}_{i=1,\cdots,N_{\rm w}}$
under three kinds of constraints, i.e, the orthonormality condition,
\begin{equation}
\label{eq:ortho_cbhfb} % -*- Eq : orthogonality relations of HFB-N (line 2)
\langle \psi_i | \psi_j \rangle =
\sum_s \int \psi_i^{\ast} (\vct{r},s) \psi_j (\vct{r},s) \dvol
 =  \delta_{ij} ,
\end{equation}
a condition on the expectation value of the number of nucleons,
\begin{equation}
\label{eq:num_cbhfb} % -*- Eq : nucleon number of HFB-N
2 \sum_{i=1}^{N_{\rm w}} v_i^2 = N,
\end{equation}
and the normalization of the $u$-$v$ factors, $ u_i^2+v_i^2=1$.

Reinhard \etal wrote that the advantage of the representation
(\ref{eq:vac_cbhfb}) over (\ref{eq:vac_hfbq}) is that one has to
consider only a single set of wave functions
$\{ \psi_i \}$, 
not a double set
$\{\phi_i, \varphi_i \}$ \cite{RBR97}.
In our opinion, however, one may expect much greater benefit from
the canonical-basis representation. Namely, $i$ may be truncated as $i
\leq N_{\rm w}$ = ${\cal O}(N)$ $\ll$ $\frac{1}{2}N_{\rm B}$ 
to a very good approximation.

The reason is the localization of the density.
The density distribution of HFB solutions can be shown to behave 
asymptotically for large $r$ as
\begin{equation}
\label{eq:density_tail} % -*- Eq : Asymptotic form of the density tail
\rho(r) \sim  \left( \frac{e^{-\kappa r}}{r} \right)^2 , \;\;\;
\kappa \ge \frac{\sqrt{-2m\epsilon_{\rm F}}}{\hbar},
\end{equation}
as far as the Fermi level $\fermilevel$ is negative \cite{DFT84},
where $m$ is the nucleon mass.
Consequently, $\psi_i$ appearing on the
right-hand side of Eq.~(\ref{eq:dens_cbhfb}) must be a localized
function as $\rho(\vct{r})$ on the left-hand side.
On the other hand,
the orthogonality relation (\ref{eq:ortho_cbhfb}) restricts the number of
wave functions which can exist in the vicinity of the nucleus.
Therefore the number of canonical orbitals cannot be very large.
In mesh representations, 
$N_{\rm B}$ is proportional to the volume of the box 
while $N_{\rm w}$ is proportional to the volume of the nucleus.
The latter is $10^{-1} - 10^{-2}$ times as small as the former
in typical calculations.

Situation is quite different in the quasiparticle formalism: The number
of quasiparticle states is proportional to the volume of the box. 
This is because the localization of the density demands only that of
$\varphi_i$ through Eq.~(\ref{eq:dens_hfbq}) while $\phi_i$ does not
have to be localized in general.
The orthogonality condition (\ref{eq:norm_qp}) involves both $\varphi_i$
and $\phi_i$.  This enables many quasiparticle states having similar
$\varphi_i$ to be orthogonal to each other by differing their $\phi_i$.

%=============================================================================
\section{Mean fields for zero-range interactions \label{sec:meanfields}}

% << HFB with density-dependent zero-range forces : Interactions >>

In this paper, we employ the Skyrme interaction \cite{Sk56,VB72},
which is a density- and momentum-dependent zero-range interaction:
\begin{eqnarray}
\hat{v}  & = & t_0 (1+x_0 P_{\sigma}) \delta
+ {\textstyle \frac{1}{2}} t_1 (1+x_1 P_{\sigma}) 
(\vct{k}^2 \delta + \delta \vct{k}^2) \nonumber \\
& + & t_2 (1+x_2 P_{\sigma}) \vct{k} \cdot \delta \vct{k}
 + {\textstyle \frac{1}{6}} t_3 (1+x_3 P_{\sigma}) \rho^{\alpha} \delta ,
\label{eq:mf_int} % -*- Eq : Interaction for the mean-field channel
\end{eqnarray}
where $P_{\sigma}$ is the spin exchange operator,
$\delta = \delta \left( \vct{r}_1 -\vct{r}_2 \right)$
with $\vct{r}_1$ and $\vct{r}_2$ the position vectors of the interacting 
two nucleons, 
$\rho$ is the nucleon density at $\vct{r}_1$ ($=\vct{r}_2$), and
\begin{equation}
\label{eq:relativemomentum} % -*- Eq: relative momentum
\vct{k} = \frac{1}{2i} \left( 
\nabla_{\vct{r}_1} -\nabla_{\vct{r}_2}
% \frac{\partial}{\partial \vct{r}_1} -\frac{\partial}{\partial \vct{r}_2} 
\right).
\end{equation}
$\hbar \vct{k}$ is the relative momentum between the two nucleons.
Since it is hermitian under the box (vanishing and periodic) 
boundary conditions,
we have not specified in which way (to bra or ket states) it operates 
in Eq.~(\ref{eq:mf_int}) unlike in, e.g., Ref.~\cite{VB72}.
Zero-rangeness makes the mean-field potentials local,
which is an essential advantage for coordinate-space solutions.

Among the terms of the Skyrme force,
% on the right-hand size of Eq.~(\ref{eq:mf_int}),
those with strength parameters $t_0$, $t_1$, and $t_3$ act only on
s-wave relative motions.
The term with $t_1$ serves to take into account the effects of the
finite-rangeness in terms of the lowest order momentum dependence, 
while the term with $t_3$ expresses a density dependence.
% (not treated as a three-body force including when $\alpha$=1)
%
The term with strength $t_2$ acts on relative p-wave states
through a different form of momentum dependence. 
Note that the isospin channel ($T$=0 or 1)
is uniquely determined through the requirement
of the two-body antisymmetry.
Owing to the p-wave as well as the s-wave interaction terms,
the Skyrme force can act on 
all the four kinds of spin-isospin channels.

The complete Skyrme interaction includes a spin-orbit term
$i W (\vct{\sigma_1} + \vct{\sigma_2}) \cdot \vct{k} \times \delta \vct{k}$,
which we have excluded from Eq.~(\ref{eq:mf_int}).
This elimination decreases the size of the numerical calculations by a
factor of four (two from spin up and down components, another two from
real and imaginary parts of the spatial wave function).
Due to this size reduction, we can perform very severe tests of the
canonical-basis HFB method by, e.g., 
taking into account unusually many canonical orbitals or 
expressing wave functions on a very large mesh.

As the parameters of the force, 
we choose the SIII set \cite{BFG75}
except for the spin-orbit term, i.e.,
$t_0$ = $-1128.75$ MeV fm$^3$,
$t_3$ =$14000$ MeV fm$^{3+3\alpha}$,
$\alpha=1$, and $W=0$.
As for the parameters appearing in the momentum dependent terms, 
the following two combinations are sufficient to treat $N=Z$ even-even nuclei,
\begin{eqnarray}
f_1 & = & {\textstyle \frac{1}{16}}\left(  3 t_1 + 5 t_2 + 4 t_2 x_2 \right),\\
f_2 & = & {\textstyle \frac{1}{64}}\left( -9 t_1 + 5 t_2 + 4 t_2 x_2 \right),
\end{eqnarray}
whose values are
$f_1$ =     44.375 MeV fm$^5$ and $f_2$ =   $-62.969$ MeV fm$^5$.

As is usually done, we employ different parameters between the mean-field
and the pairing interactions.
We use the following interaction for the pairing channel,
\begin{eqnarray}
\hat{v}_{\rm p} & = & v_{\rm p} \frac{1-P_{\sigma}}{2} 
\left[
  \left\{ 
      1 - \frac{\rho}{\rho_{\rm c}}
      - \left( \frac{\tilde{\rho}}{\tilde{\rho}_{\rm c}}\right)^2
  \right\} \delta 
\right.  \nonumber \\
  & & \left.
  - \frac{1}{2 k_{\rm c}^2} \left( \vct{k}^2 \delta + \delta \vct{k}^2 \right)
\right] ,
\label{eq:pair_int} % -*- pairing interaction
\end{eqnarray}
where $v_{\rm p}$ is the strength and
$\frac{1}{2}(1-P_{\sigma})$ is the projector into spin-singlet states.
Terms in the braces represent dependences on particle and
pairing densities.
The interaction becomes repulsive where $\rho > \rho_{\rm c}$ or 
$\tilde{\rho} > \tilde{\rho}_{\rm c}$.
For the particle-density dependence, 
we use $\rho_{\rm c}$ = 0.32 fm$^{-3}$
to vanish the volume-changing effect\cite{TBF93}
(This choice, twice of the matter density,
is also recommended for a different reason \cite{DNS01}.)
The dependence on the pairing density $\tilde{\rho}$
(to be defined by Eq.~(\ref{eq:pairingdensity})) has been introduced 
in Ref.~\cite{Taj99a} to prevent unphysical behaviors 
of high-lying canonical orbitals discussed
in Appendix \ref{sec:pointcollapse}. 
This term is squared because $\tilde{\rho}$ can become negative in principle.
Except in Appendix \ref{sec:pointcollapse},
we use $\tilde{\rho}_{\rm c} = \infty$.
The momentum dependent term acts on s-wave relative motions
and quenches interactions between nucleons in high-momentum states.
The interaction vanishes at relative momentum $k_{\rm c}$ 
at low-density points.
If $\tilde{\rho_{\rm c}} = k_{\rm c} = \infty$,  this
pairing force is reduced to that introduced in \cite{TBF93}.
The typical values of the parameters are $v_{\rm p}=-1000$ MeV fm$^3$ and
$k_{\rm c}$=2 fm$^{-1}$ but different values are also used.

The $S$=0 and $T$=1 part of the Skyrme force of Eq.~(\ref{eq:mf_int}) 
may be rewritten as Eq.~(\ref{eq:pair_int}) without the
pairing density dependent term.
Using different parameterizations between the two forces
has an advantage of avoiding confusions.

% << Hamiltonian density for even-even N=Z nuclei >>

When one considers both protons and neutrons, the state of the nucleus
is usually assumed to be a product of two BCS-type states of
Eq.~(\ref{eq:vac_cbhfb}), one for the protons and the other for the
neutrons:
\begin{equation}
\label{eq:vac_pn} % -*- Eq : total (protons + neutrons) wave function
| \Psi \rangle  = \prod_{\rm \porn=p}^{\rm n} 
\prod_{i=1}^{N_{\rm w}^{(\porn)}}
\left( 
  u_{\porn i} + v_{\porn i} 
  \; a^{\dagger}_{\porn i} \; a^{\dagger}_{\porn \bar{\imath}}
\right) | 0 \rangle,
\end{equation}
where the index q distinguishes between protons (p) and neutrons (n).
$a^{\dagger}_{{\rm p} i}$ creates a proton having a wave function
$\psi_{{\rm p} i}(\vct{r},s)$ while
$a^{\dagger}_{{\rm n} i}$ creates a neutron with a wave function
$\psi_{{\rm n} i}(\vct{r},s)$.
This product form matches our choice of the pairing interaction
of Eq.~(\ref{eq:pair_int}) which acts only on $T=1$ pairs.

In this paper we treat only $N$=$Z$ nuclei without Coulomb interaction
for the sake of simplicity.
In this case, the wave functions are the same between protons and neutrons.
Moreover, because the potentials are independent of the spin, the
wave functions $\psi_i(\vct{r},s)$ can be factorized into a product
of a spin wave function and a real-number function of the position,
which we write $\psi_i(\vct{r})$ hereafter.
More specifically, we assume
  $\psi_{{\rm p} i}(\vct{r},s)$ 
= $\psi_{{\rm n} i}(\vct{r},s)$ 
= $\psi_{i}        (\vct{r}) \delta_{s,\frac{1}{2}}$  and
  $\psi_{{\rm p}\bar{\imath}}(\vct{r},s)$
= $\psi_{{\rm n}\bar{\imath}}(\vct{r},s)$
= $\psi_{i}        (\vct{r}) \delta_{s,-\frac{1}{2}}$.

The isoscalar pairing may take over the isovector pairing 
in $N \sim Z$ nuclei \cite{Go79}. 
Nevertheless, we assume Eq.~(\ref{eq:vac_pn})
because the aim of this paper is to examine a method whose most important
applications are to the nuclei near the neutron drip line.

Our Hamiltonian $H$ consists of a kinetic energy term and
two-body interaction terms. 
It should be understood that both are expressed in the second quantized 
form because HFB states do not have a fixed number of particles.
The kinetic energy term is $-\hbar^2 \nabla^2 / 2 m$ in 
the one-body expression, for which we assume the mass of a nucleon $m$
to be the average of the proton and the neutron bare masses 
($m_{\rm p}$ and $m_{\rm n}$) with
an approximate correction factor for the center-of-mass motion
of a nucleus of mass number $A$:
\begin{equation}
\label{eq:nucleonmass} % -*- Eq : mass of a nucleon corrected for CM motion
m = \left(1-\frac{1}{A} \right)^{-1} \frac{m_{\rm p} + m_{\rm n}}{2}.
\end{equation}
For the interaction terms,
we use the interaction (\ref{eq:mf_int}) for the mean-field type contractions
and the interaction (\ref{eq:pair_int}) for the pairing type contractions
in evaluating the matrix elements of the two-body interactions
using Wick's theorem.

The total energy for the state (\ref{eq:vac_pn}) can be expressed in terms of
a single space integral as,
\begin{equation}
\label{eq:tot_eng} % -*- Eq : total energy of HFB vacuum
  E_{\rm tot}
= \langle \Psi \vert H \vert \Psi \rangle
= \int {\cal H}(\vct{r}) \dvol ,
\end{equation}
%
% where the integrand is called the Hamiltonian density and given by
where the integrand is given by
\begin{eqnarray}
{\cal H} & = &
\frac{\hbar^2}{2m} \tau + \frac{3}{8}t_0 \rho^2 
+ f_1 \rho \tau
+ f_2 \rho \nabla^2 \rho
+\frac{1}{16}t_3 \rho^{2+\alpha} \nonumber \\
&+ & \mkern-9mu \frac{v_{\rm p}}{8} \mkern-6mu
\left[
  \left\{
    \mkern-2mu 1 \mkern-2mu 
    - \mkern-2mu \frac{\rho}{\rho_{\rm c}} \mkern-2mu
    - \mkern-2mu \left( \frac{\tilde{\rho}}{\tilde{\rho}_{\rm c}}\right)^2
  \right\} \tilde{\rho}^2
  -\frac{\tilde{\rho}}{k_{\rm c}^2}
  \left( \tilde{\tau}-\nabla^2 \tilde{\rho} \right)
\right] .
\label{eq:ham_dens} % -*- Eq : Hamiltonian density (line 2)
\end{eqnarray}
On the right-hand side of the above equation, functions of
position $\vct{r}$ are
\begin{eqnarray}
\label{eq:particledensity} % -*- Eq : rho
\rho (\vct{r}) & = &
4 \sum_{i=1}^{N_{\rm w}} v_i^2 | \psi_i (\vct{r}) |^2, \\
\label{eq:kineticenergydensity} % -*- Eq : tau
\tau (\vct{r}) & = &
4 \sum_{i=1}^{N_{\rm w}} v_i^2 | \vct{\nabla} \psi_i (\vct{r}) |^2,\\
\label{eq:pairingdensity} % -*- Eq : rho tilde
\tilde{\rho}(\vct{r}) & = &
4 \sum_{i=1}^{N_{\rm w}} u_i v_i | \psi_i (\vct{r}) |^2 ,\\
\label{eq:pairingkineticenergydensity} % -*- Eq : tau tilde
\tilde{\tau} (\vct{r}) & = &
4 \sum_{i=1}^{N_{\rm w}} u_i v_i | \vct{\nabla} \psi_i (\vct{r}) |^2 ,
\end{eqnarray}
which are called \cite{DFT84}
the (particle) density,
the kinetic-energy density,
the pairing density, and
the pairing kinetic-energy density,
respectively.

There is a set of single-particle operators 
(${\cal H}_i$, $h$, and $\tilde{h}$)
which are useful to obtain mean-field solutions.
They can be obtained by taking the functional derivative of $E_{\rm tot}$
with respect to a canonical wave function $\psi_i$. 
This turns out to be equivalent to
operating a state-dependent single-particle Hamiltonian ${\cal H}_i$ to
the wave function :
\begin{equation}
\label{eq:functionalderivativeofetot} % -*- Eq : H_i psi_i
  \frac{\delta E_{\rm tot}}{\delta \psi_i}
= 4 {\cal H}_i \psi_i^{\ast} , \;\;\;
  \frac{\delta E_{\rm tot}}{\delta \psi_i^{\ast}}
= 4 {\cal H}_i \psi_i .
\end{equation}
Here, we have taken the familiar point of view that 
$\psi_i$ and $\psi_i^{\ast}$ can be formally regarded as independent variables,
although $\psi_i$ is actually a real function.

One can express ${\cal H}_i$, in turn, as
a linear combination of state-independent single-particle Hamiltonians as,
\begin{equation}
\label{eq:statedependenthamiltonian} % -*- Eq : H_i
{\cal H}_i  =  v_i^2 h + u_i v_i \tilde{h},
\end{equation}
where $h$ and $\tilde{h}$ are called the mean-field and pairing Hamiltonians. 
They are given by
\begin{eqnarray}
\label{eq:ham_mf} % -*- Eq. : mean-field Hamiltonian
h & = & - \vct{\nabla} \cdot B \vct{\nabla} + V ,\\
\label{eq:ham_pair} % -*- Eq : pairing Hamiltonian (=potential)
\tilde{h} & =& - \vct{\nabla} \cdot \tilde{B} \vct{\nabla} + \tilde{V} ,
\end{eqnarray}
with
\begin{eqnarray}
\label{eq:mf_B} % -*- Eq. : mean-field B
B  & = & \frac{\hbar^2}{2m} + f_1 \rho,  \\
\label{eq:mf_pot} % -*- Eq. : mean-field potential
V  & = & \mkern-5mu \frac{3}{4} t_0 \rho 
+ {\textstyle \frac{2+\alpha}{16}} t_3 \rho^{1+\alpha} 
+ f_1 \tau + 2 f_2 \nabla^2 \rho
- \frac{v_{\rm p} \tilde{\rho}^2 }{8 \rho_{\rm c}} , \\
\label{eq:pair_B} % -*- Eq. : pairing B
\tilde{B}  & =  &
- \frac{v_{\rm p}}{8 k_{\rm c}^2} \tilde{\rho}, \\
\label{eq:pair_pot} % -*- Eq : pairing potential
\tilde{V} & = & \mkern-9mu \frac{v_{\rm p} }{4}  \mkern-6mu
\left[
  \left\{ 1 - \frac{\rho}{\rho_{\rm c}} 
         - 2 \left(
                       \frac{\tilde{\rho}}{\tilde{\rho}_{\rm c}} 
             \right)^2
  \right\} \tilde{\rho}
  -\frac{\tilde{\tau}}{2 k_{\rm c}^2}
  +\frac{\nabla^2 \tilde{\rho}}{k_{\rm c}^2}
\right] .
\end{eqnarray}
We call $V$ and $\tilde{V}$ the mean-field and 
the pairing potentials, respectively.

When there are no pairing correlations,
the mean-field Hamiltonian $h$ is identical to the 
HF single-particle Hamiltonian.
Therefore, its expectation value,
\begin{equation}
\label{eq:sp_eng} % -*- Eq : single-particle energy
\epsilon_i = \langle \psi_i | h | \psi_i \rangle,
\end{equation}
may be called the single-particle energy of the $i$th canonical orbital.
The negative of the expectation value of the pairing Hamiltonian,
\begin{equation}
\label{eq:statedepgap} % -*- Eq : state dependent pairing gap
\Delta_i = - \langle \psi_i | \tilde{h} | \psi_i \rangle,
\end{equation}
has a meaning of the pairing gap of the orbital.
The pairing gap should be positive or zero.
Indeed,
$\Delta_i$ could take negative values only if the wave function 
would consist of very high-momentum components \cite{TOT94}.
A related discussion is given in Sec.~\ref{sec:canorb}.

The quasiparticle states of Eq.~(\ref{eq:ani_op_qp})
are the eigenstates of their own Hamiltonian, i.e.,
so-called HFB super matrix composed of $h$ and $\tilde{h}$:
\begin{equation}
\label{eq:hfb_matrix} % -*- Eq. : HFB super matrix
\left( \begin{array}{cc} h - \fermilevel & \tilde{h} \\ 
\tilde{h} & - h + \fermilevel \end{array} \right)
\left( \begin{array}{c} \phi_i \\ \varphi_i \end{array} \right)
= E_{\rm qp}^{(i)}
\left( \begin{array}{c} \phi_i \\ \varphi_i \end{array} \right),
\end{equation}
where $\fermilevel$ is the Fermi level.
In small subspaces like several major shells,
the easiest way to obtain an HFB solution is to solve
Eq.~(\ref{eq:hfb_matrix}) as an eigenvalue problem of a hermitian matrix.
In large spaces like mesh representations,  however,
it is easier to obtain the canonical orbitals directly,
rather than via quasiparticle states.
Such a direct method is explained in the next section.

Incidentally, 
if one considers seriously that $\psi_i$ is a real function, 
one may notice that the functional derivative should be doubled since
\begin{equation}
\label{eq:commentfuncderiv} % -*- Eq: comment on functional derivative 
  \frac{\delta E_{\rm tot}}{\delta \psi_i}
= \frac{\delta E_{\rm tot}}{\delta \psi_i^{\rm bra}}
              \frac{\delta \psi_i^{\rm bra}}{\delta \psi_i}
+ \frac{\delta E_{\rm tot}}{\delta \psi_i^{\rm ket}}
              \frac{\delta \psi_i^{\rm ket}}{\delta \psi_i}
= 2 \cdot 4 {\cal H}_i \psi_i ,
\end{equation}
as $\psi_i^{\rm bra} = \psi_i$ and $\psi_i^{\rm ket} = \psi_i$.
Eq.~(\ref{eq:commentfuncderiv}) seems to be a more appropriate
expression of what our numerical calculation program actually does.
However, the factor $2$ at the beginning of the right-hand side as well as
the factor $4$ which reflects the spin-isospin degeneracy do not matter
because they can be conveniently absorbed into the step-size parameter 
of the gradient method (see Sec.~\ref{sec:gradientmethod}).

%=============================================================================
\section{Cut-off schemes for the pairing interaction \label{sec:cutoff}}

In this section we discuss on the cut-off schemes for zero-range pairing
interactions in relation to the canonical-basis HFB method.

% << 1 >>  Delta function force needs cut-off.

Delta function forces without a cut-off lead to a divergence of the
strength of the pairing correlation.
In order to prevent the divergence,
one usually takes only those
quasiparticles whose excitation energy $E_{\rm qp}^{(i)}$ is lower
than some cut-off energy $\eqpcut$ in constructing
the ground state \cite{DFT84}.
Namely Eq.~(\ref{eq:vac_hfbq}) is modified to
\begin{equation}
\label{eq:vac_hfbq_cut}
\vert \Psi \rangle = \prod_{i=1}^{N_{\rm B}}
\theta \left( \eqpcut - E_{\rm qp}^{(i)} \right) b_i |0\rangle,
\end{equation}
where $\theta$ is the Heaviside (step) function.

% << 2 >> We regard cut-off as one of the constants to define the interaction.

In this paper we do not discuss how one should change the strength as
a function of $E_{\rm c}$ for renormalization.  Instead, we regard
$E_{\rm c}$ as one of the constants to define the force.

% << 3 >> Summary of the discussion in this section

In order to implement the cut-off,
one should pay attention to the following two peculiarities of 
the canonical-basis HFB method.
i) The introduction of $E_{\rm c}$ causes a serious inconvenience 
to this method, i.e., the absence of the quasiparticle Hamiltonian. 
For this reason, it is preferable to use 
the number of canonical orbitals $N_{\rm w}$ instead of $E_{\rm c}$.
ii) For delta-function pairing interactions,
substitution of $E_{\rm c}$ with $N_{\rm w}$ leads to
a physically meaningless situation, 
in which canonical orbitals are collapsed to a spatial point
(a point collapse phenomenon, hereafter).
To overcome this difficulty, we modify the pairing interaction by adding
a repulsive momentum-dependent term.
In the rest of this section we discuss these two problems.

% << 4 >> N_W have to be used instead of E_c for delta function forces

i) Let us explain how the introduction of the cut-off causes
the absence of the quasiparticle Hamiltonian.
An analogy with the BCS approximation indicates
a natural way to introduce an energy cut-off 
to the canonical-basis method.
Namely, the smooth cut-off method\cite{BFH85} 
modifies the seniority interaction as
\begin{equation}
\label{eq:cut_sen_force} % -*- Eq. : seniority force with smooth cut-off
\hat{v}_{\rm pair}  =  -G
\left( \sum_{i>0} \; f_i \; a^{\dagger}_i a^{\dagger}_{\bar{\imath}} \right)
\left( \sum_{j>0} \; f_j \; a_{\bar{\jmath}} a_j \right),
\end{equation}
where $f_i = f(\epsilon_i)$ is a function of single-particle energy
$\epsilon_i$ and takes on $\sim 1$ well below a
chosen cut-off energy and smoothly becomes zero above it.
The smoothness is indispensable to the stability of the solution.
In analogy, we may modify the pairing density as
\begin{equation}
\label{eq:cut_pair_dens} % -*- Eq. : cut-off pairing density
\tilde{\rho} = 4 \sum_{i=1}^{N_{\rm w}} \; u_i v_i \; \vert \psi_i \vert^2
\;\; \rightarrow \;\;
4 \sum_{i=1}^{N_{\rm w}} \; f_i \; u_i v_i \; \vert \psi_i \vert^2
\end{equation}
with
\begin{equation}
\label{eq:cut_func_k} % -*- Eq. : cut-off function of momentum
f_i = \exp \left( -\frac{\mu^2}{4} k_i^2 \right), \;\;
k_i^2 = - \int \psi_i^{\ast}(\vct{r}) \nabla^2 \psi_i(\vct{r}) \dvol.
\end{equation}
In Eq.~(\ref{eq:cut_func_k}), we assume a dependence on the
kinetic energy ($\propto k_i^2$), not on the mean-field energy
$\epsilon_i$ as in Eq.~(\ref{eq:cut_sen_force}), to avoid a highly
complicated expression of the gradient for the latter case. (In BCS,
such complications are simply neglected.)
The gradient is given for $k_{\rm c}=\infty$ by
\begin{equation}
\label{eq:grad_cutoff} % -*- Eq : gradient with cut-off
\frac{1}{4} \frac{\delta E_{\rm tot}}{\delta \psi_i^{\ast}} =
\left[  v_i^2 h + u_i v_i \tilde{h} \left\{ 
f(k_i^2) + f'(k_i^2) \nabla^2 \right\} \right] \psi_i .
\end{equation}
This method works well to prevent the point collapses in numerical tests
using $\mu = 1.2$ fm taken from the Gogny force \cite{DG80}.
However, this cut-off scheme has an disadvantage that the HFB super
matrix in Eq.~(\ref{eq:hfb_matrix}) cannot be defined,
because in Eq.~(\ref{eq:grad_cutoff}) the pairing Hamiltonian (the part
proportional to $u_i v_i$ in square brackets) is dependent on the
canonical-orbital index $i$.
This means that there is no common
pairing Hamiltonian to all the quasiparticles.  Consequently one does
not have well-defined Bogoliubov quasiparticle states and cannot utilize
them to refine the HFB solution, which is a rather serious drawback as
discussed in the next section.
If one can use $N_{\rm w}$ in place of $E_{\rm c}$, however, 
one can recover a unique quasiparticle Hamiltonian.

Incidentally, a similar kind of ambiguity arises from
the cut-off scheme in terms of the quasiparticle energies
according to Eq.~(\ref{eq:vac_hfbq_cut}). In this case,
the canonical orbitals do not have well-defined Hamiltonians.
It may not matter, however, because one does not need Hamiltonians
to define canonical orbitals when they can be obtained by transforming 
the quasiparticle states.

% << 5 >>  Then, point collapse occurs.

ii) Let us discuss the phenomenon of the point collapse.
In quasiparticle (or two-basis) HFB method,
specifying the number of quasiparticle (HF single-particle) states 
to be considered has the same consequences as
imposing an energy cut-off
because the states to be chosen are anyway the eigenstates
of the lowest energies of the quasiparticle (HF) Hamiltonian.

On the other hand, the canonical-basis HFB method is very peculiar 
concerning this point.
The canonical orbitals are not the eigenstates of a single
operator but each of them has its own Hamiltonian ${\cal H}_i$
as a function of the occupation probability given by
Eq.~(\ref{eq:statedependenthamiltonian}).
${\cal H}_i$ becomes $h$ and $\tilde{h}$ 
in the two extreme situations $v_i^2$=1 and 0, respectively.
It is only when one considers a small number of orbitals
that all of them are the lowest-energy approximate eigenstates of $h$.
Otherwise,
some of the orbitals may become the eigenstates of $\tilde{h}$
rather than $h$:
We show in Appendix \ref{sec:pointcollapse} that,
if the occupation probability of an orbital becomes sufficiently small,
it spontaneously decreases the probability further toward zero 
(to decrease the total energy of the nucleus)
and change its Hamiltonian to $\tilde{h}$.
This occurs when  the pairing force is a delta function 
(with or without particle-density dependence),
for which $\tilde{h}$ is a local potential and thus its eigenstates are
delta functions in the spatial coordinate. 
This means that their $\langle h \rangle$ is infinitely high.
On the other hand, the rest of the orbitals are left in
low-energy subspace of $h$.
%
% The intermediate-energy states between the two groups are missing. 
%
Therefore the restriction on $N_{\rm w}$ 
results in utterly different solutions from
those obtained by using a cut-off energy explicitly.

This peculiarity of the canonical-basis method is essential to
the implementation of the method.
In the quasiparticle and the two-basis methods,
the unphysical divergence of the pairing correlation strength
occur only in the limit of $N_{\rm w} \rightarrow \infty$ 
(supposing $N_{\rm w}$ is used instead of $E_{\rm c}$).
In the canonical-basis method, in addition to this divergence,
a different type of unphysical behavior (the point collapse)
can also happen for $N_{\rm w}$ fixed at finite numbers.
One has to take special care of the latter problem 
if one tries to implement the canonical-basis method.

% << 6 >> Then, one needs momentum dependence.

Now, if one can suppress the point collapse,
one is allowed to use $N_{\rm w}$ as the control parameter of 
the cut-off energy.
Appendix \ref{sec:pointcollapse} shows that
this can be achieved, e.g., by introducing a sufficiently strong
repulsive momentum dependence to the delta-function pairing interaction.
With this term applied to finite (small or large) numbers of canonical
orbitals, unphysical situations including the point collapse never happen.
(If it is applied to infinite number of orbitals,
not the point collapse but the divergence occurs \cite{TOT94}.)
In this paper we employ this momentum-dependent force.

In addition to the role to enable a cut-off in terms of $N_{\rm w}$,
equally important is the byproduct that it
also clarifies the nature of high-lying canonical
orbitals by providing a kinetic energy term to the pairing
Hamiltonian $\tilde{h}$.
Further explanations using a numerical example
are given in Sec.~\ref{sec:canorb}.

Incidentally, Appendix \ref{sec:pointcollapse} also gives a discussion on
an alternative way in terms of a pairing-density dependent force
to enable cut-off in terms of $N_{\rm w}$.

% << 7 >> Referee said we should discuss on regularization and renormalization.

Before ending this section, we mention the prospect about
the renormalization of the pairing force strength.
Considering the recent quantitative success of the regularization method
of delta-function forces based on the Thomas-Fermi approximation
\cite{BY02,Bul02,YB03,GU03},
we think it is worth trying in the canonical-basis method in future. 

However, it will require a special treatment.
The Thomas-Fermi approximation makes the 
interaction strength a decreasing function of the particle density.
One might expect this modification of the interaction would hinder
the point collapse. Actually, dependences on particle density cannot
prevent it because the collapse makes only the pairing density 
divergent but leaves the particle density unchanged
(because $v \propto w^{3/2}$ and $|\psi_{\rm w}(0)|^2 \propto w^{-3}$
as shown in Appendix \ref{sec:pointcollapse}).
Therefore one has to keep the momentum-dependent term.
This will lead to a modification of the regularization procedure. 

% Although a sharp cut-off is possible for momentum-dependent interactions, 
% it will be better if one can establish a relation between $E_{\rm c}$ to
% be used in the renormalization and $N_{\rm w}$ used (instead of 
% $E_{\rm c}$) to obtain the HFB solutions.

%==============================================================================
\section{Gradient method for canonical-basis HFB  \label{sec:gradientmethod} }

We explain in this section the details of a procedure to
obtain the canonical-basis solution of the HFB equation
directly, not by way of quasiparticle states.
What should be done is to
minimize $E_{\rm tot}$ of Eq.~(\ref{eq:tot_eng})
under constraints of Eqs.~(\ref{eq:ortho_cbhfb})
and (\ref{eq:num_cbhfb}).
Equivalently, one may introduce a Routhian,
\begin{eqnarray}
\nonumber           % -*- Eq : Routhian of HFB-N (line 1)
R & = & E_{\rm tot} - 4 \fermilevel \sum_{i=1}^{N_{\rm w}} v_i^2 \\
\label{eq:routhian} % -*- Eq : Routhian of HFB-N (line 2)
& - & 4 \sum_{i=1}^{N_{\rm w}} \sum_{j=1}^{N_{\rm w}} \lambda_{ij} \left\{
\langle \psi_i | \psi_j \rangle - \delta_{ij} \right\},
\end{eqnarray}
and minimize it without constraints.
$\epsilon_{\rm F}$ is probably the most familiar Lagrange multiplier,
whose physical meaning is the Fermi level.
In the definition (\ref{eq:routhian}), $N_{\rm w}^2$ Lagrange
multipliers $\lambda_{ij}$ obeying hermiticity,
\begin{equation}
\label{eq:hermiticity} % -*- Eq : requirement of hermiticity of lambda_{ij}
\lambda_{ij} = \lambda_{ji}^{\ast},
\end{equation}
are introduced instead of $\frac{1}{2}N_{\rm w}(N_{\rm w}+1)$ 
independent multipliers.
The hermiticity ensures the equality between the number of constraints and the
number of independent multipliers.
Moreover, it makes $R$ real
so that two conditions, $\delta R / \delta \psi_i =0$ and $\delta R /
\delta \psi_i^{\ast} =0$, become equivalent and thus one has to
consider only one of them.
Note that $\delta_{ij}$ is subtracted from
$\langle \psi_i | \psi_j \rangle$,
in contrast to Ref.~\cite{RBR97}.
This subtraction is necessary to treat $\lambda_{ij}$ not as constants like
$\epsilon_{\rm F}$ but as functionals of the wave functions.

The stationary condition of $R$ results in two kinds of equations.
One is $\partial R / \partial v_i =0$, which concerns the
occupation amplitudes $v_i$ and is fulfilled by \cite{RBR97}
\begin{equation}
\label{eq:bcs} % -*- Eq : BCS-like equation to determine v_i
v_i^2 = \frac{1}{2} \pm \frac{1}{2} \frac{\epsilon_i -\epsilon_{\rm F} }
{\sqrt{(\epsilon_i- \epsilon_{\rm F} )^2+\Delta_i^2}}.
\end{equation}
Among the double sign, the minus sign corresponds to the minimum.
The relative sign between $v_i$ and $u_i$ = $\pm\sqrt{1-v_i^2}$ 
should be plus, i.e., $u_i v_i \ge 0$ (unless $\Delta_i < 0$).
The stationary condition for a wave function $\psi_i(\vct{r})$
leads to the following equation:
\begin{eqnarray}
\nonumber % -*- Eq : (d R)/(d psi) =0 (line 1)
\frac{1}{4} \frac{\delta R}{\delta \psi_i^{\ast}} & = &
{\cal H}_i \psi_i - \sum_{j=1}^{N_{\rm w}} \lambda_{ij} \psi_j \\
\label{eq:gradient} % -*- Eq : (d R)/(d psi) =0 (line 2)
& - &
\sum_{j=1}^{N_{\rm w}} \sum_{k=1}^{N_{\rm w}} 
\frac{\delta \lambda_{jk}}{\delta \psi_i^{\ast}}
\left\{ \langle \psi_j | \psi_k \rangle - \delta_{jk} \right\} = 0,
\end{eqnarray}

Owing to the state dependence of ${\cal H}_i$, the orthogonality
condition becomes nontrivial to fulfill.  In HF, the orthogonality is
automatically satisfied because $\psi_i$ are eigenstates of the same
hermite operator $h$, which are orthogonal to one another.  The
orthogonalization procedure is needed only because of the instability
for decaying into Pauli-forbidden configurations.  On the other hand, in
the canonical-basis HFB method, the orthogonalization is indispensable
and the explicit functional form of $\lambda_{ij}$ is the most
important secret of the method.

Reinhard \etal have proposed \cite{RBR97}
\begin{equation}
\label{eq:lambda1} % -*- Eq : Reinhard's lambda_{ij}
\lambda_{ij} = \frac{1}{2}
\langle \psi_j | \left( {\cal H}_i + {\cal H}_j \right) | \psi_i \rangle.
\end{equation}
Let us give our reasoning on how the above form can be deduced.
This consideration plays the crucial role in order to modify the
form for the sake of a faster convergence.
The requirement that Eq.~(\ref{eq:gradient}) must
hold at the solution (where $\langle \psi_i | \psi_j \rangle = \delta_{ij}$)
means,
\begin{equation}
\label{eq:lambda1b} % -*- Eq : Required form of lambda_{ij} at the solution
{\cal H}_i | \psi_i \rangle = \sum_{k=1}^{N_{\rm w}} \lambda_{ik}
| \psi_k \rangle 
\;\;\; (\mbox{at the solution}).
\end{equation}
By taking the overlap with $\langle \psi_j |$, one obtains,
\begin{equation}
\label{eq:lambda2} % -*- Eq : Required form of lambda_{ij} at the solution
\lambda_{ij} = \langle \psi_j | {\cal H}_i | \psi_i \rangle
\;\;\; (\mbox{at the solution}).
\end{equation}
Eqs.~(\ref{eq:lambda1}) and (\ref{eq:lambda2}) are equivalent at the
solution because $\lambda_{ij}$ is defined to be hermite
by Eq.~(\ref{eq:hermiticity}).
However, one should employ
Eq.~(\ref{eq:lambda1}) rather than Eq.~(\ref{eq:lambda2})
because the former, but not the latter, is hermite at points other than
the solution.

% << Gradient method >>

One can utilize the gradient method
to obtain the HFB solutions in the canonical-basis formalism.
Let us describe this method briefly:
Small variations in $\psi_i$ and $\psi_i^{\ast}$ change the Routhian as,
\begin{equation}
\label{eq:deltarouthian} % -*- Eq : delta Routhian
\delta R  = \frac{\delta R}{\delta \psi_i} \delta \psi_i
          + \frac{\delta R}{\delta \psi_i^{\ast}} \delta \psi_i^{\ast},
\end{equation}
which indicates the steepest descent direction to be,
\begin{equation}
\label{eq:deltapsi} % -*- Eq : delta psi
\delta \psi_i \propto -\frac{\delta R}{\delta \psi_i^{\ast}} , \;\;\;
\delta \psi_i^{\ast} \propto -\frac{\delta R}{\delta \psi_i}.
\end{equation}
One takes a small distance movement 
toward this direction, i.e.,
$\psi_i \rightarrow \psi_i + \delta \psi_i$ with
\begin{equation}
\label{eq:gradstep} % -*- Eq : a gradient step
\delta \psi_i = - \itss 
\left\{
  {\cal H}_i \psi_i - \sum_{j=1}^{N_{\rm w}} \lambda_{ij} \psi_j
\right\} .
\end{equation}
By repeating the evaluation of $\{ \delta \psi_i \}$ and the movement,
one eventually reaches the minimum of $R$.
Incidentally,
in the braces on the right-hand side, 
the term including $\delta \lambda_{jk}/\delta \psi_i^{\ast}$
in the second member of Eqs.~(\ref{eq:gradient}) is dropped because 
we know empirically that the orthogonality is stable without this term.

In Eq.~(\ref{eq:gradstep}),
$\itss$ is a parameter introduced to control the size of a movement.
One should regard that the prefactors on the right-hand side of
Eq.~(\ref{eq:commentfuncderiv}) are included in this parameter.
One may call $\itss$ the imaginary-time-step size because
the first order approximation in $\itss$ of
the imaginary-time evolution method \cite{DFK80}
becomes essentially the same procedure
as the gradient method :
\begin{eqnarray}
\delta \psi_i & = & e^{-\itss {\cal H}_i} \psi_i -\psi_i \nonumber \\ 
& = & -\itss {\cal H}_i \psi_i + {\cal O}\left(\itss^2 \right).
\label{eq:imag_evo} % -*- Eq : imaginary-time evolution
\end{eqnarray}
%

% << requirement on the imaginary-time-step size >>

In order to justify the negligence of the higher-order terms 
in Eq.~(\ref{eq:imag_evo}), it must hold\cite{TTO96}
\begin{equation}
\label{eq:delta_tau} % -*- Eq : imaginary time step size
f_{\rm evo} < 2 \;\;\; \mbox{for} \;\;\;
\itss = \frac{f_{\rm evo}}{\ekinmax} ,
\end{equation}
where $\ekinmax$ is the maximum kinetic energy
(used as the upper bound of the single-particle energy).
For the Cartesian mesh representation with a mesh spacing $a$, 
\begin{equation}
\label{eq:ekinmax} % -*- Eq : maximum kinetic energy
\ekinmax = 3 \left( \frac{\pi}{a} \right)^2 B(\rho) ,
\end{equation}
where $B(\rho)$ is given by Eq.~(\ref{eq:mf_B}).
By choosing $\rho$ between $\rho = 0$ (free space) and
$\rho$=0.16 fm$^{-3}$ (nuclear matter density),
one obtains $\ekinmax$ = 1254 MeV for the SIII force with $a=0.8$ fm$^{-1}$.
For the calculations shown in this paper, we have used
$f_{\rm evo}=1.0$, i.e., $\itss = 5 \times 10^{-25}$ sec.
Kinetic energy also arises from the pairing Hamiltonian $\tilde{h}$.
However, an estimation by replacing $B(\rho)$ of Eq.~(\ref{eq:ekinmax}) with 
$\tilde{B}(\tilde{\rho})$ of Eq.~(\ref{eq:pair_B}) results in
an order of magnitude smaller value.

% << Acceleration of the convergence >>

What is presented above may be called a na\"{\i}ve version of the
gradient method.  The convergence to the HFB solution turns out to be
very slow with this na\"{\i}ve method. A numerical example will be given
in Sec.~\ref{sec:conv}.  The origin of this slowness can be traced back
to the factor $u_i v_i$ in ${\cal H}_i$, which is just a linear
combination of $h$ and $\tilde{h}$ with coefficients $v_i^2$ and $u_i
v_i$.  The effects of ${\cal H}_i$ can be very weak for canonical
orbitals having small $v_i$,
% ($\epsilon_i$ $-$ $\epsilon_{\rm F}$ $\gg$ $\Delta_i$)
which leads to the smallness of their changes in a gradient-method step.

Now, steepest-descent paths depend on the choice of the variables.
For example, Eq.~(\ref{eq:gradstep}) is obtained when one uses
$\psi_i$ as the variables.
If one uses norm-resized wave functions 
$\chi_i = \psi_i / \sqrt{\alpha_i} $, where $\alpha_i$ is a scaling factor,
a gradient-method step becomes 
$\delta \chi_i \propto - \delta R / \delta \chi_i^{\ast}$,
which is equivalent to
\begin{equation}
\label{eq:grad_psi} % -*- Eq : gradient method for w.f. psi after scaling
\delta \psi_i = - \itss \alpha_i \frac{\delta R}{\delta \psi_i^{\ast}}
= - \alpha_i \itss
\left( {\cal H}_i \psi_i - \sum_j \lambda_{ij} \psi_j \right) .
\end{equation}
Thus the path is changed.
By choosing $\alpha_i$ to cancel the small factors in ${\cal H}_i$,
one can accelerate the otherwise slow convergence,
which we call the accelerated gradient method.

In this paper, we take
\begin{equation}
\label{eq:alphai} % -*- Eq : alpha_i
\alpha_i \simeq {\rm min}\, \left( \frac{1}{v_i^2}, 
\frac{f_{\rm acc}}{u_i v_i} \right),
\end{equation}
where the factor $f_{\rm acc}$ may be chosen empirically 
to maximize the speed of convergence.
We choose $f_{\rm acc}=5$.
This factor seems to take care of the fact that
$\tilde{h}$ is smaller than $h$ by an order of magnitude.
With this choice for the scaling factor,
$\alpha_i {\cal H}_i \sim h$ for deeply bound orbitals
and $\alpha_i {\cal H}_i \sim 5 \tilde{h}$
for high-lying orbitals. 
Thus, all the orbitals evolve at roughly the same pace.

Incidentally, by using the approximate equality symbol in Eq.~(\ref{eq:alphai}),
we mean the utilization of some common recipes for numerical
calculations like taking a moving average and restricting on the maximum
values.

% << Improved lambda_{ij} >>

It should be noticed here that
one should also modify the form of multipliers $\lambda_{ij}$ 
when one introduces the acceleration factors $\alpha_i \not= 1$.
In principle one may reach the solution by using Eq.~(\ref{eq:lambda1}) 
because transformations of variables do not change the location
of the minimum of $R$.
Nevertheless, one should modify Eq.~(\ref{eq:lambda1}) for a practical reason.
In calculating the gradient vector whose exact form is given by the 
second member of Eqs.~(\ref{eq:gradient}),
the last term takes much more computing time than the first two terms
due to $\delta \lambda_{jk} / \delta \psi_i^{\ast}$.  One can drop
the last term if the orthogonality relation (\ref{eq:ortho_cbhfb}) is
fulfilled along the path of the evolution.  Let's suppose that
the relation is satisfied before a gradient-method step is taken and
require that it is conserved without the dropped term
to the first order in $\itss$ after the step, i.e.,
\begin{equation}
\label{eq:orthog_path} % -*- Eq : conservation of orthogonality along path
\langle \psi_i | \psi_j \rangle = \delta_{ij},
\;\;\;
\langle \psi_i' | \psi_j' \rangle = \delta_{ij} +{\cal O}
\left( \left( \itss \right)^2 \right) ,
\end{equation}
with $\psi_i'= \psi_i + \delta \psi_i$ 
and $\delta \psi_i$ given by Eq.~(\ref{eq:grad_psi}) .
Substituting $\psi_i$ and $\psi_i'$ in Eq.~(\ref{eq:orthog_path})
and requiring the hermiticity (\ref{eq:hermiticity})
result in
\begin{equation}
\label{eq:lambda3} % -*- Eq : lambda_{ij} for acceleration factor != 1
\lambda_{ij}= \frac{1}{\alpha_i+\alpha_j}
\langle \psi_j | \left( \alpha_i {\cal H}_i + \alpha_j {\cal H}_j \right)
| \psi_i \rangle.
\end{equation}
This new form of $\lambda_{ij}$, as well as the original form of 
Eq.~(\ref{eq:lambda1}), 
fulfills the requirement that it should
agree with the expression (\ref{eq:lambda2}) at the solution.
The original form differs from the new form, however,
before reaching the solution if $\alpha_i \not= \alpha_j$.
Consequently, only the new form conserves the orthogonality
during the course of the evolution.

We have indeed suffered from large errors of orthogonality by using
the original form.
On the other hand, by using the new form,
we have observed that the error does not grow
but decreases without performing explicit orthogonalizations.
The reason for this stability will probably be found in the second order 
terms in $\itss$ neglected in Eq.~(\ref{eq:orthog_path}).

% << Diagonalization of HFB super matrix >>

There is an auxiliary method to speedup the convergence.
It is to diagonalize the HFB super matrix given by Eq.~(\ref{eq:hfb_matrix})
in the subspace spanned by $N_{\rm w}$ canonical orbitals.
Then one obtains $N_{\rm w}$ quasiparticle states with positive energy,
whose two-component wave functions defined in Eq.~(\ref{eq:ani_op_qp}) are
expressed as,
\begin{equation}
\label{eq:qpwf_in_cbsp} % -*- Eq : qp wavefunction in canonical-basis subspace
\phi_i (\vct{r}) = \sum_{j=1}^{N_{\rm w}} \psi_j (\vct{r}) U_{ji} , \;\;\;  
\varphi_i (\vct{r}) = \sum_{j=1}^{N_{\rm w}} \psi_j (\vct{r}) V_{ji} ,
\end{equation}
where $(U_{1i},\cdots,U_{N_{\rm w}i},V_{1i},\cdots,V_{N_{\rm w}i})^T$ 
is the $i$th normalized eigenvector of Eq.~(\ref{eq:hfb_matrix}).
The HFB ground state is constructed as the vacuum of these quasiparticles
as in Eq.~(\ref{eq:vac_hfbq}).
By diagonalizing the one-body density matrix of this vacuum
in this subspace, $V^{\ast} V^{T}$, 
one obtains a renewed set of canonical orbitals as the
eigenvectors and $v_i^2$ as the eigenvalues.
If the initial set of canonical orbitals are taken from the
exact solution, the renewed and the initial sets are identical.
However, if the initial set is taken before the state converges to the
solution, the renewed set corresponds to a better solution than the
initial set.
It is because the above procedure gives the exact variational
minimum \cite{Go79,RS80} in the subspace 
spanned by the initial canonical orbitals.
The gradient-step method takes care of both the variation inside this
subspace and the optimization of the subspace itself. 
The diagonalization performs only the former part but 
perfectly in a single step.
This diagonalization method is not indispensable but useful to
obtain the solutions. It makes the convergence more robust
and somewhat quicker if it is performed after every $\sim$100
gradient-method steps. 
Its effect seems to saturate with this interval since 
using shorter periods does not lead to noticeable improvements.
A numerical example is given in Sec.~\ref{sec:conv}.

Incidentally,
the idea of this diagonalization method originates in the 
two-basis formalism of HFB \cite{GBD94}, in which
the quasiparticle Hamiltonian is diagonalized in the subspace spanned
by low-lying HF orbitals (eigenstates of $h$), not the canonical
orbitals as in this paper.

The relation between these quasiparticles and the true quasiparticles
defined in the full space is discussed in Sec.~\ref{sec:cbqp}.

%-------------------------------------------------------------------------
\section{Results of numerical calculations \label{sec:results}}

% << Default set up of the calculations >>

In this section, we show the results of numerical calculations
using a newly developed computer program \cite{Taj98a,Taj98b,Taj99a,Taj00a} 
of the canonical-basis HFB method 
in a three-dimensional Cartesian mesh representation.
The parameters of the model are as follows:
For the mean-field interaction, Skyrme SIII force is employed but
its spin-orbit term and the Coulomb force are not included.
For the pairing interaction,
$\rho_{\rm c}=0.32$ fm$^{-3}$ and $\tilde{\rho}_{\rm c}=\infty$ are used.
The value of $v_{\rm p}$ is changed depending 
on $k_{\rm c}$ (= 1.7 -- 5 fm$^{-1}$) and $N_{\rm w}$ (= 14 -- 210) to make
the average pairing gap to be 2 -- 2.5 MeV.
Unless the values are specified,
$k_{\rm c}$=2 fm$^{-1}$, $v_{\rm p}=-1050$ MeV fm$^3$, and $N_{\rm w}=21$
are used.
With $N_{\rm w}=21$, the number of orbitals amounts to three times as large
as the number of the nucleons, since each orbital can be
occupied by four nucleons due to the spin-isospin degeneracy.

The calculated state is the axially symmetric prolate solution of 
$^{28}_{14}$Si$_{14}$.
The single-particle wave functions are expressed on a Cartesian mesh
having (in the default set up)
$29 \times 29 \times 29$ mesh points with mesh spacing $a$=0.8 fm.  
As the boundary condition, we assume a cubic box 
whose infinitely high walls are located at the 0th and the 30th mesh points.  
The center of mass of the nucleus is placed at the center of the box.
The 17-point formulae \cite{Taj01a} are employed to calculate the first and second
derivatives.  In applying these formulae, wave functions are assumed to
be antisymmetrically extended beyond the walls.
Spatial integrals are done using the mid-point (=trapezoidal) rule.

%---------- Convergence to HFB solutions (job262) ----------------------------
\subsection{Convergence to HFB solutions \label{sec:conv}}

\begin{figure}%G:psf
\includegraphics[width=8.6 cm]{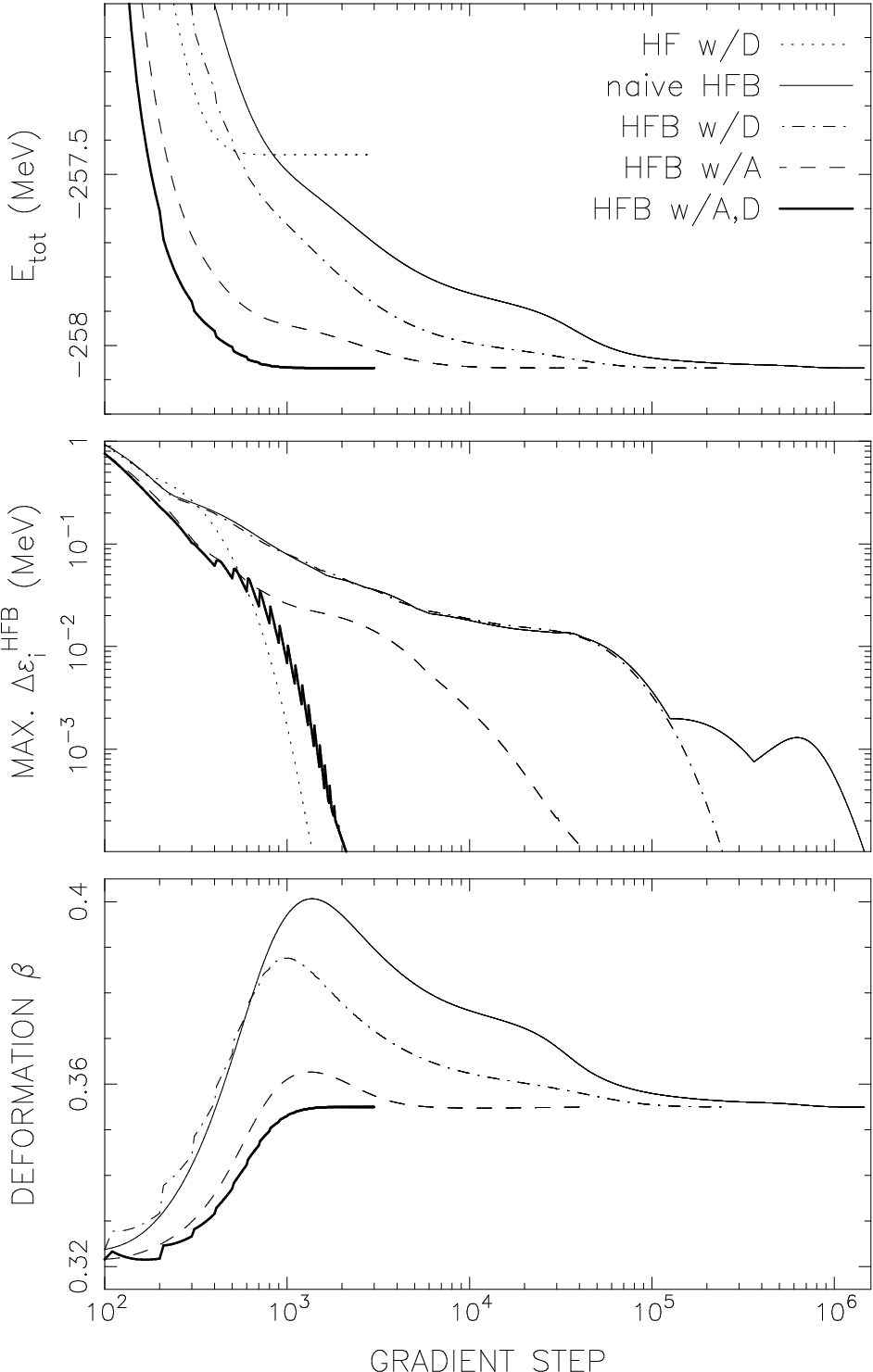}%G:psf
\caption{Convergence to HF and HFB solutions for $^{28}$Si.
See text for explanations.
}
\label{fig:conv}
\end{figure}

In Fig.~\ref{fig:conv}, examples of the convergence curves are shown
for several versions of the gradient method.
The calculations are done using the default parameters.
The initial wave functions of the canonical orbitals are taken 
from the eigenstates of the standard harmonic oscillator potential
with a quadruple deformation of $\beta = 0.3$ and $\gamma = 15^{\circ}$.
The $D_{2h}$ symmetry is conserved to a very good accuracy during the course
of the gradient-method evolutions.

For the size of an imaginary time step, we use  $f_{\rm evo}=1.0$, defined by
Eq.~(\ref{eq:delta_tau}), for all the curves.  This value
of $f_{\rm evo}$ roughly optimizes the convergence speed for the fastest
method.  For slower methods, $f_{\rm evo}$ may be
slightly larger but such a small change does not affect our discussions
below.

In connection with the step size, we also adopt a recipe from the HF+BCS
method of Ref.~\cite{BFH85}, according to which the changes of $h$ and
$\tilde{h}$ after each gradient-method step should be further damped compared
with the changes of $\psi_i$ (by a factor of 0.4 in our case).  It is a
precaution against occasional instabilities like oscillating behaviors.
This additional damping does not affect our conclusion on the comparison
of the methods, either.

%-------------------------- F I G U R E -----------------------------------
%\begin{figure}[htb]%G:box
%\begin{center}\framebox{Figure \ref{fig:conv}}\end{center}%G:box
%------------------- E N D   O F   F I G U R E ----------------------------

The top portion of Fig.~\ref{fig:conv}
shows the total energy $E_{\rm tot}$ of Eq.~(\ref{eq:tot_eng}).
The middle portion shows
the maximum over $i=1,\cdots,N_{\rm w}$
of the error of the second equality of Eqs.~(\ref{eq:gradient})
(neglecting the contribution from the error of orthogonality),
\begin{equation}
\Delta \epsilon_i^{\rm HFB} = 
  \biggl\vert 
    {\cal H}_i \psi_i - \sum_{j=1}^{N_{\rm w}} \lambda_{ij} \psi_j 
  \biggr\vert ,
\end{equation}
which is an indicator of the accuracy of HFB solutions.
As for HF, this quantity is reduced to the maximum over 
$i=1,\cdots,\frac{1}{4}A$ of
\begin{equation}
\Delta \epsilon_i^{\rm HF} = 
  \left\vert 
    h \psi_i - \epsilon_i \psi_i 
  \right\vert
= \sqrt{
    \langle \psi_i \vert h^2 \vert \psi_i \rangle - 
    \langle \psi_i \vert h   \vert \psi_i \rangle^2
  },
\end{equation}
which is just the energy width of $\vert \psi_i \rangle$ for $h$.
The bottom portion shows the mass quadrupole deformation parameter $\beta$
defined as the general definition \cite{Go79}
times $\sqrt{45/16\pi} \simeq 0.95$ disregarding nucleon form factors.
HF result for $\beta$, converging to 0.44, is omitted.
The abscissae are common to all the portions and designate the number of
gradient-method steps.

Thin solid lines in all the three portions are the result of the
na\"{\i}ve (i.e., with $\alpha_i=1$) gradient method.
These curves demonstrate that one can indeed obtain an HFB solution
with the canonical-basis HFB method on a mesh,
because after a million steps, $E_{\rm tot}$ and $\beta$ appear to reach
a plateau and $\Delta \epsilon_i^{\rm HFB}$ becomes as small as 0.1 keV.
We have also obtained similar convergence curves for other quantities
like the axial asymmetry $\gamma$ and the r.m.s.\ radius.
There is a serious problem, however, 
of the obvious slowness of the convergence.

Dot curves show the convergence to an HF solution 
with pairing interactions turned off.  
(With ``w/D'' in the legend in the top portion of the figure we mean
that diagonalization of $h$ is done in the subspace of
occupied $\frac{1}{4}A = 7$ HF orbitals to renew the orbitals after
every 100 gradient-method steps.
Without this diagonalization, $\psi_i$ are not HF orbitals 
and do not satisfy $\Delta \epsilon_i^{\rm HF}=0$.
This diagonalization also makes the convergence somewhat quicker.)
One can see that the HF energy reaches a plateau by three orders of magnitude
faster than the na\"{\i}ve HFB method.
In the middle portion, while HF can achieve precision of 0.1 keV with
only 1400 steps, HFB requires 55000 steps for 10 keV, $3\times 10^5$
steps for 1 keV, and $1.5 \times 10^6$ steps for 0.1 keV precisions.

A method to improve the convergence speed is 
the diagonalization of the HFB super matrix
in the subspace spanned by $N_{\rm w}$ canonical orbitals.
Dot-dash curves are obtained by performing this
diagonalization after every 100 steps.
One can see that the convergence speed is improved by an order of magnitude.
However, this progress is not satisfactory yet compared with the HF curve.

Dash curves uses the accelerated gradient
method (i.e., $\alpha_i \not= 1$, expressed as ``w/A'' in the legend), which
brings about a speedup by two orders of magnitude.
A comparison with dot-dash curves suggests that 
the variation of the canonical-basis subspace
itself is a more important factor than a minimization inside
the subspace.

Using both the diagonalization and the acceleration methods leads to
thick solid curves (expressed as ``w/A,D'' in the legend).
Steep changes in every 100 steps are due to the diagonalizations.
One can see that these curves converge almost as quickly as
the dot curves of the HF method.
These results demonstrate that canonical-basis HFB can be solved
without very heavy numerical computations by adopting the
two improvements described above.

%-------- Properties of canonical orbitals (job285) -----------------------

\subsection{Properties of the canonical-basis orbitals \label{sec:canorb}}

\begin{figure*}%G:psf
\includegraphics[width=17.8 cm]{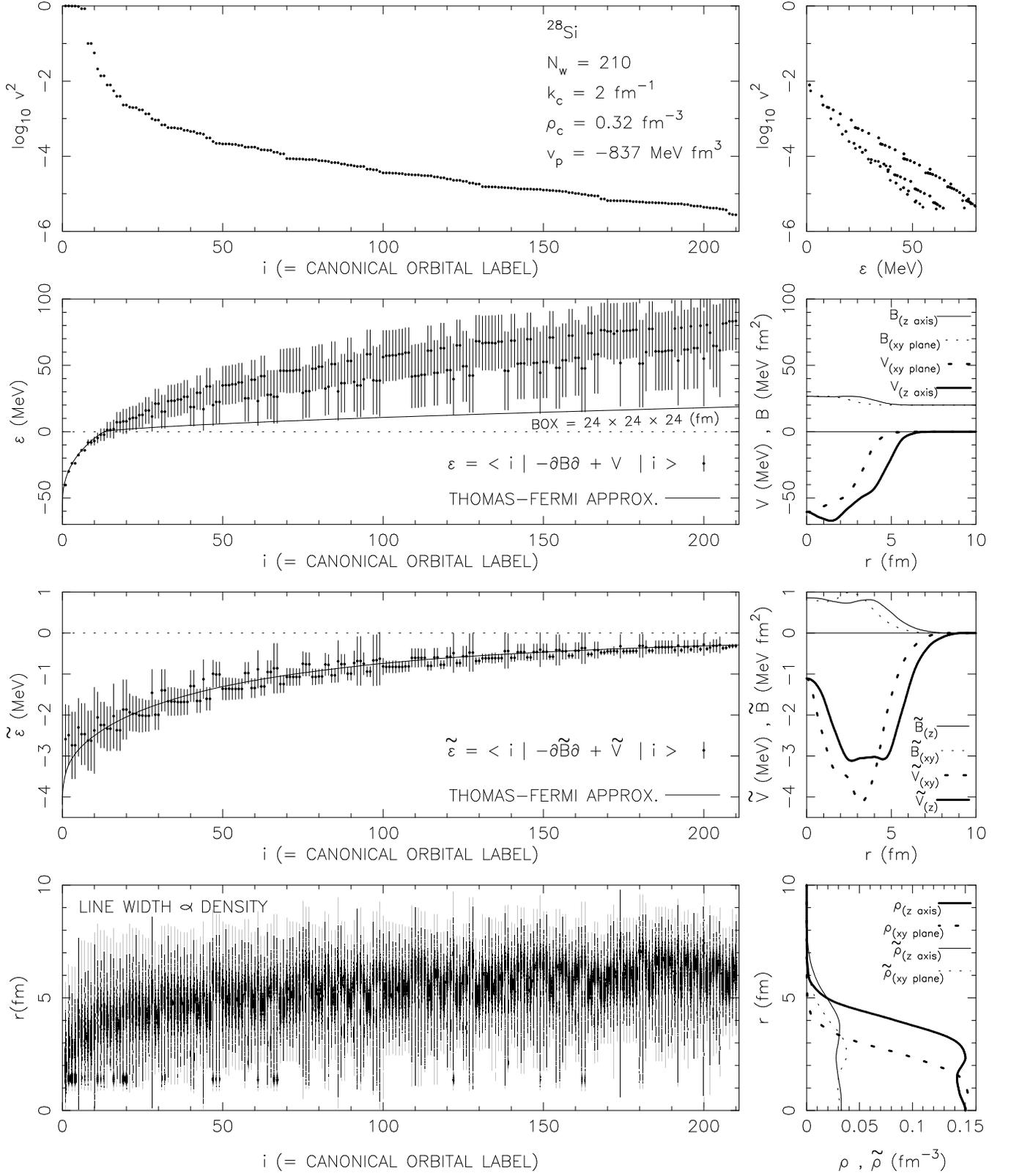}%G:psf
\caption{
Relations between the properties of canonical-basis orbitals and the
mean-field and pairing Hamiltonians for $^{28}$Si.
Panels in the left column show
occupation probability, mean-field energy, pairing energy, and radial
density distribution for each canonical-basis orbital.
Panels in the right column show the occupation probability
(again, but versus the mean-field energy), mass and potential terms of the
mean-field Hamiltonian, those of pairing Hamiltonian, and the
density distribution of the nucleus.
See text for explanations.
}
\label{fig:canorb}
%\end{figure}%G:box
\end{figure*}

In this subsection, we use Fig.~\ref{fig:canorb} to discuss the
properties of the canonical orbitals in connection with the mean-field and
the pairing Hamiltonians.
An analysis using these two Hamiltonians leads to a comprehensive
understanding of the nature of high- as well as low-lying canonical
orbitals.
The calculation has been done for $^{28}$Si with $N_{\rm w}=210$
canonical orbitals,
which is 30 times as many as the number of the orbitals below the Fermi level.
For the combination of this $N_{\rm w}$ and $k_{\rm c}=2$ fm$^{-1}$,
we have adjusted the strength of the pairing force to be $v_{\rm p}=-837$ MeV
so that the average pairing gap have a reasonable size, 2.0 MeV.

The result shown in Fig.~\ref{fig:canorb} 
has been obtained after $1.2 \times 10^5$ gradient-method steps:
Because of the slow convergence of very high-lying orbitals
(even using the accelerated method),
we continued the gradient-method evolution until no changes are
going on in any of $\epsilon_i$.
The last change took place during $5\times 10^4$th to $8 \times 10^4$th
steps, in which two orbitals gradually increased their energies
from $\sim 50$ MeV to $\sim 80$ MeV. During this change
$\Delta \epsilon_i^{\rm HFB}$ increased slightly up to $3 \times 10^{-5}$ MeV.
In the last $4 \times 10^4$ steps there were no changes and
$\Delta \epsilon_i^{\rm HFB}$ was decreased to $< 10^{-5}$ MeV. 

%-------------------------- F I G U R E -----------------------------------
%\begin{figure}[htb]%G:box
%\begin{center}\framebox{Figure \ref{fig:canorb}}\end{center}%G:box
%G:psf
%------------------- E N D   O F   F I G U R E ----------------------------

The top-left panel shows 
the logarithm of the occupation probability $v_i^2$.
The abscissa (common to all the four panels in the left column)
is the label $i$ of canonical orbitals,
which are sorted in the descending order of $v_i^2$. 
First seven orbitals have occupation probabilities close to unity.
Above them, the decrease of the probability looks rather steady.
It is in fact a result of a mixing of several sequences:
The top-right panel plots the same $v_i^2$ data versus $\epsilon_i$
(only $\epsilon_i > 0$ part is shown),
in which one can see three sequences. 
Let us denote the number of the oscillator quanta by $\nosc$.
We have confirmed that the top sequence corresponds to $\nosc =l$ subshells 
($4 \le l \le 9$ are completely seen in $\epsilon >0$ region).
The lower bunch of dots are for $l \le \nosc -2$.
For $\epsilon > 40$ MeV,
$l=\nosc -2$ sequence is bifurcated from $l \le \nosc -4$ sequences.
These observations indicate that the canonical orbitals have the same
shell structure as that of the harmonic oscillator at least up to
several tens of MeV.
Then, a question arises what generates this shell structure.
The second and third top portions give clues to answer this question.

The panels in the second top row show the mean-field Hamiltonian $h$
and its expectation value and width for each
canonical orbital. In the right panel, the mass
term $B$ and the potential term $V$ of $h$ are plotted as functions of
the distance from the center of the nucleus. 
Their profiles in the symmetry ($z$)
axis as well as those in the equatorial ($xy$) plane are shown.
In the left panel, mean-field energy of
each canonical orbital is designated using a dot and a vertical line.
The dot corresponds to the expectation value $\epsilon_i$ while
the vertical line connects between $\epsilon_i \pm \Delta \epsilon_i$ with
\begin{equation}
\label{eq:delta_epsilon} % -*- Eq : Delta epsilon_i
\epsilon_i = \langle \psi_i | h | \psi_i \rangle, \;\;\;
\Delta \epsilon_i = \sqrt{\langle \psi_i | h^2 | \psi_i \rangle 
-\epsilon_i^2}.
\end{equation}
One can see that first seven orbitals, which are below the Fermi level, 
have very narrow widths (30 -- 760 keV): 
They are approximate eigenstates of
$h$, i.e., the (occupied) HF orbitals.  Above the Fermi level, the width
becomes larger with increasing expectation value and thus the
wave functions of canonical orbitals begin to diverge from those of
(unoccupied) HF orbitals. 
At the crossing point from negative to positive $\epsilon_i$,
there seems to be only a smooth continuation of this trend.
This is an essential difference between the HFB canonical and the HF orbitals.
Concerning the HF orbitals, 
$\epsilon_i$ may be estimated in the Thomas-Fermi approximation, where
the number of levels of $h$ below $\epsilon$ is given as,
\begin{equation}
\label{eq:thomas_fermi} % -*- Eq : Thomas-Fermi approx to the level density.
\Gamma(\epsilon) = \frac{1}{6\pi^2}\int 
\left\{ \frac{\epsilon- V(\vct{r})}{B(\vct{r})} \right\}^{3/2} 
\theta \bigl( \epsilon- V(\vct{r}) \bigr)
\dvol .
\end{equation}
The widths of HF orbitals are always zero.
The solid curve designates the function $i=\Gamma(\epsilon)$.
It is owing to the finite volume of the normalization box 
that $\Gamma(\epsilon)$ is not infinite for positive $\epsilon$. 
One can see a distinct difference between this curve for the HF orbitals
and the dots for the HFB canonical orbitals: 
The level density is by far sparser in the latter.
(It is indeed the reason to use canonical orbitals 
as discussed in Sec.~\ref{sec:cbhfb}.)
It seems difficult to relate the level structure of canonical orbitals to $h$.

Incidentally, one can see a bifurcation of the sequence of the dots
($i$, $\epsilon_i$) at $i \sim 30$. Its origin is the subshell structure:
Upper sequence corresponds to high-angular momentum ($l=\nosc$) subshells
while the lower one to $l \le \nosc -2$ subshells.

The third panels from the top show similar graphs to the second top panels
but for the pairing Hamiltonian $\tilde{h}$.
In the right panel, attention should be paid to the scale of the ordinate,
which is smaller by an oder of magnitude than that for $h$.
It is because the expectation value of $\tilde{h}$ is nothing
but the negative pairing gap (i.e., $\tilde{\epsilon}_i = -\Delta_i$)
and thus its size is only a few MeV.
In the left panel, dots are plotted at $\tilde{\epsilon}_i$ and
vertical lines connect between
$\tilde{\epsilon}_i \pm \Delta \tilde{\epsilon}_i$, where
\begin{equation}
\label{eq:delta_epsilon_tilde} % -*- Eq : Delta epsilon_tilde_i
\tilde{\epsilon}_i = \langle \psi_i | \tilde{h} | \psi_i \rangle, \;\;\;
\Delta \tilde{\epsilon_i} = \sqrt{\langle \psi_i | \tilde{h}^2 | \psi_i\rangle 
-\tilde{\epsilon}_i^2}.
\end{equation}

One can see an opposite trend of the width between $h$ and $\tilde{h}$:
In the former (latter), the width is smaller for lower (higher)
expectation values.
It suggests a view that high-lying canonical orbitals are closely related to
$\tilde{h}$.
This can be understood in terms of the state dependent Hamiltonian,
${\cal H}_i = v_i^2 h + u_i v_i \tilde{h}$.
Because of this form, 
it is close to $h$ ($\tilde{h}$) for deeply bound (highly excited)
canonical orbitals.

The solid curve designates the Thomas-Fermi estimation of 
$\tilde{\Gamma}(\tilde{\epsilon})$,
the number of levels of $\tilde{h}$ below $\tilde{\epsilon}$.
$\tilde{\Gamma}$ can be obtained by replacing
$\Gamma, \epsilon, V, B $ with
$\tilde{\Gamma}, \tilde{\epsilon}, \tilde{V}, \tilde{B} $,
respectively, in Eq.~(\ref{eq:thomas_fermi}).
One can see that this curve agrees quite well with the dots.
(Sorting the orbitals more appropriately in the ascending order of
$\tilde{\epsilon}_i$ makes the agreement better.)
It supports the view that $\tilde{h}$ roughly determines the
level structure of canonical orbitals.
It also guarantees that the application of the Thomas-Fermi approximation 
to $\tilde{h}$ provides a rather precise method to count
the number of canonical orbitals.

One can see another difference between $h$ and $\tilde{h}$
that $\epsilon_i$ becomes positive for $i > 15$ 
while $\tilde{\epsilon}_i$ is still negative at $i = 210$.
It means that $\tilde{h}$ has by far more number of bound states than $h$. 
Its origins in the Thomas-Fermi approximation are
the larger ratio $\tilde{V}/\tilde{B}$ than $V/B$ inside the nucleus and 
the vanishing behavior of $\tilde{B}$ as $r \rightarrow \infty$.

It is an interesting question whether there are finite or infinite
number of canonical orbitals with $v_i^2 >0$. 
Non-zero occupation probability means $\Delta_i >0$ 
and thus $\tilde{\epsilon}_i < 0$.
Therefore one has to count the number of orbitals with 
negative $\tilde{\epsilon}$.
To give a Thomas-Fermi estimation of this number,
let us suppose the form of density tails to be,
\begin{equation}
\label{eq:rho_tau_tail} % -*- Eq : rho and tau tails
\tilde{\rho} = \tilde{\rho}_{\rm in} \frac{e^{-\kappa r}}{\kappa r}, \;\;\;
\tilde{\tau} = \tilde{\tau}_{\rm in} \frac{e^{-\kappa r}}{\kappa r}.
\end{equation}
For the kinetic-energy density inside the nucleus, a reasonable estimation is
$\tilde{\tau}_{\rm in} = k_{\rm F}^2 \tilde{\rho}_{\rm in}$,
where $\hbar k_{\rm F}$ is the Fermi momentum.
Using a relation $\nabla^2 \tilde{\rho} = \kappa^2 \tilde{\rho}$
and assuming $\rho \ll \rho_{\rm c}$ appropriate in the peripheral region,
one can show that $\tilde{V}(r)/\tilde{B}(r)$ is independent of $r$.
Consequently, the number of orbitals becomes proportional to the
volume of the integral region $V_{\rm box}$. 
The explicit form of the result is
\begin{equation}
\label{eq:gammazero} % -*- Eq : gamma_0
\Gamma(\tilde{\epsilon}=0) = V_{\rm box} \Gamma_0, \;\;\;
\Gamma_0 = \frac{1}{6\pi^2}\left( 2k_{\rm c}^2 - 2 \kappa^2 - k_{\rm F}^2
\right).
\end{equation}
Therefore, the number of non-zero-occupation canonical orbitals is infinite.

The bottom panels show 
the radial distributions of the nucleon density (right panel)
and that of each canonical orbital (left panel).
The former is shown in two directions 
while the latter is integrated over the angles as,
\begin{equation}
\rho_i (r) =  r^2 \int \vert \psi_i (\bm{r}) \vert d\hat{r},
\end{equation}
and thus normalized as 
$
\int_0^{\infty} \rho_i (r) dr = 1
$.
Vertical lines are drawn in the interval of $r$ 
where $\rho_i(r) > 0.001$ fm$^{-1}$.
The width of the lines are proportional to $\rho_i(r)$, except that
the width is saturated at 0.8 (in the scale of the abscissa) for
$\rho_i(r) \ge 0.5$ fm$^{-1}$.
One can see that the canonical orbitals are localized in the
neighborhood of the nucleus and only gradually shifted outward with
increasing $i$.
It is not due to the finite box size because
even the 210th orbital is located far from the boundary, which is
at 12 to 12$\sqrt{3}$ fm.
The location of the maximum of $\rho_i(r)$ agrees fairly well with
the classical turning point of $\tilde{V}$ for energy $\tilde{\epsilon}_i$
(as seen in the third top panel in the right column).
This vindicates the view that
$\tilde{h}$ determines the spatial extent of high-lying canonical orbitals.

Let us give another fact to confirm this view.
Evaluating Eq.(\ref{eq:gammazero}) using
$k_{\rm c}=2$ fm$^{-1}$, $\kappa=1$ fm$^{-1}$ (see next subsection), and 
$k_{\rm F}=1.4$ fm$^{-1}$, one obtains $\Gamma_0$ = 0.14 fm$^{-3}$.
For $\Gamma(\tilde{\epsilon}=0) = N_{\rm w}=210$, the value of $V_{\rm box}$
becomes equal to the volume of a sphere of radius 7.1 fm.
This roughly agrees with the fact that the 210th orbital has the maximum
density at $r=7.5$ fm.

The discussions of this subsection can be summarized into three points:
i) Canonical orbitals well below the Fermi level 
are localized by the mean-field potential. 
ii) Highly excited canonical orbitals are localized 
by the pairing potential.
iii) The pairing Hamiltonian can have infinite number of localized
orbitals due to the vanishing mass term at large radii.

It should be noted here that we do not insist that $\tilde{h}$ alone
determines high-lying orbitals to the details.
One has to keep in mind an asymmetry between low- and high-lying orbitals.
Because lower-lying orbitals have much stronger influences on the total
energy of the nucleus, the determinations of their wave functions are
hardly affected by the orthogonality conditions with the higher-lying
orbitals. However, the opposite is not true: 
High-lying orbitals must be orthogonal to the orbitals below the Fermi level, 
which are determined mainly by $h$.
Consequently, although the most essential points can be explained by the
idea of the change of the Hamiltonian from $h$ to $\tilde{h}$, there may
still remain something delicate in the treatments of high-lying canonical
orbitals compared with those of low-lying ones.  

%-------- Tails of canonical-orbital wave functions (job260) ----------------

\subsection{Tails of canonical-basis wave functions
\label{sec:wf_tail}}

Wave functions of bound states have an asymptotic form for large
radii of
\begin{equation}
\label{eq:psi_ir} % -*- Eq : asymptotic form of canonical orbitals
\vert \psi_i(\vct{r}) \vert \sim \frac{e^{-\kappa_i r}}{r}.
\end{equation}
In HF, negative energy orbitals have the form (\ref{eq:psi_ir})
with $\kappa_i$ determined by $\epsilon_i$ as
$\hbar \kappa_i = \sqrt{- 2m\epsilon_i}$.
In HFB, the hole component of quasiparticle wave functions ($\varphi$)
has this form with $\kappa$ determined by the
excitation energy $E_{\rm qp}$ as
$\hbar \kappa = \sqrt{2m ( E_{\rm qp} - \fermilevel )}$
\cite{DFT84,DNW96}.

As for the canonical orbitals of HFB, however, one cannot relate
$\kappa_i$ to any sort of energies in a simple manner due to the
complex influences of the requirement of orthogonality, which is rooted in
the lack of a common Hamiltonian to all the orbitals.
The only possible statement is that it is not larger than the smallest
value of $\kappa$ for quasiparticles, whose lower bound is given by the
Fermi level as in Eq.~(\ref{eq:density_tail}).

In this subsection, we compute $\kappa_i$ of canonical orbitals
numerically using angle-averaged density:
$|\psi_i(r)|^2 = \frac{1}{4\pi} \int |\psi_i(\bm{r})|^2 d \hat{r} $ .
In order to see the behaviors at large radii, we have used
a large box containing $57 \times 57 \times 57$ mesh points.
The edge of the box is $58a$=46.4 fm.
The nearest walls from the center of the nucleus is at a distance
of 23.2 fm and the farthest (i.e., the corners) at 40.2 fm.
For the sake of using the large box,
we restrict the number of orbitals to $N_{\rm w}=21$, for which
an adequate strength of the pairing interaction is $v_{\rm p}=-1050$ MeV.

%-------------------------- F I G U R E -----------------------------------
%\begin{figure}[htb]%G:box
%\begin{center}\framebox{Figure \ref{fig:wf_tail}}\end{center}%G:box
\begin{figure}%G:psf
\includegraphics[angle=-90, width=7.2 cm]{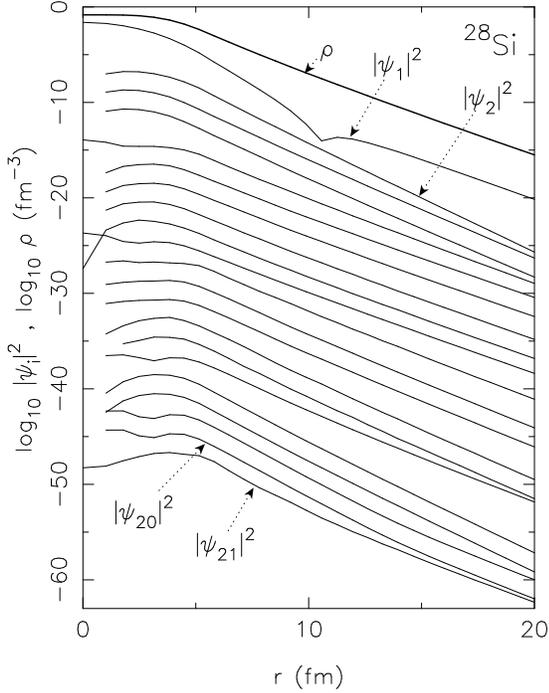}%G:psf
\caption{
Radial density profiles of canonical orbitals of $^{28}$Si.
The thick line represents the total nucleon density ($\rho$)
while the $i$th thin line from the top designates the angle-averaged
squared wave function of the $i$th canonical orbital ($|\psi_i |^2$).
The canonical orbitals are labeled from 1 to $N_{\rm w}$=21 
in the descending order of $v_i^2$.
The scale in the ordinate applies only to $\rho$ and $|\psi_1|^2$.
The other thin lines are shifted downward so that they do not overlap with 
each other.
}
\label{fig:wf_tail}
\end{figure}
%------------------- E N D   O F   F I G U R E ----------------------------

Figure~\ref{fig:wf_tail} 
shows $\vert \psi_i (r) \vert^2$ of canonical orbitals,
all of which have exponentially damping tails.
This result demonstrates that
the canonical-basis HFB method can convert
Gaussian tails of initial wave functions, for which we use eigenstates of 
a harmonic oscillator, into the correct exponential form.

It is evident that the first orbital changes its slope at around $r=10.5$ fm.
By looking at the other curves minutely, one can also find such changes
in several other orbitals.
Therefore, we give two discussions, first on the behaviors near the nucleus
and second on those far away.

%-------------------------- F I G U R E -----------------------------------
%\begin{figure}[htb]%G:box
%\begin{center}\framebox{Figure \ref{fig:wf_kappa}}\end{center}%G:box
\begin{figure}%G:psf
\includegraphics[angle=-90, width=6.5 cm]{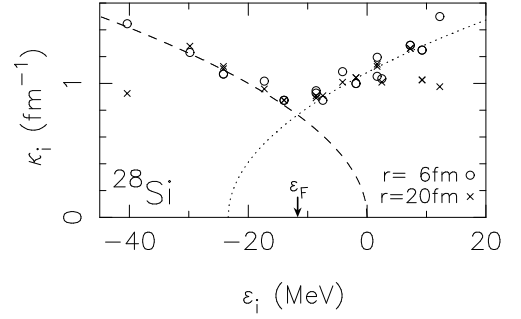}%G:psf
\caption{
Coefficient $\kappa_i$ for the tails of canonical-orbital wave functions
versus the expectation values of the mean-field Hamiltonian $\epsilon_i$.
The orbitals are the same as those shown in Fig.~\figwftail .
Open circles and crosses are for $\kappa$ at $r=6$ fm
and $r=20$ fm, respectively.
}
\label{fig:wf_kappa}
\end{figure}
%------------------- E N D   O F   F I G U R E ----------------------------

Figure~\ref{fig:wf_kappa} shows $\kappa_i$ as a function of $\epsilon_i$.
Open circles denote $\kappa_i$
determined by the logarithmic derivative of $r \vert \psi_i(r) \vert$ 
near the nucleus ($r=6$ fm).
One can see less than 21 circles because
several pairs of circles overlap exactly due to the degeneracy 
for the sign of $l_z$ ($z$-component of the orbital angular momentum).
The dash and dot curves represent functions
$\hbar \kappa_i = \sqrt{- 2m  \epsilon_i}$,
and
$\hbar \kappa_i = \sqrt{2m  (\epsilon_i - 2 \fermilevel )}$,
respectively.
The open circles seem to agree fairly well with a function
\begin{equation}
\label{eq:kappa_i} % -*- Eq : kappa_i
\hbar \kappa_i = \sqrt{2m \left(| \epsilon_i - \fermilevel | 
  - \fermilevel \right)},
\end{equation}
which coincides with the dash curve for $\epsilon_i < \fermilevel $
and the dot curve for $\epsilon_i > \fermilevel $.

By noting that the dash curve expresses $\kappa$ for bound HF orbitals,
one may understand the $\epsilon_i < \fermilevel$ part of 
Eq.~(\ref{eq:kappa_i}) as indicating that canonical
orbitals below the Fermi level have approximately the same tails 
as HF orbitals.

An alternative interpretation is also possible
which refers to the full domain of Eq.~(\ref{eq:kappa_i}):
Canonical orbitals with $\epsilon_i$ = $\fermilevel \pm E_{\rm qp}$
have the same $\kappa$ as the hole component $\varphi$ 
of a quasiparticle state with $E_{\rm qp}>0$.
It may suggest that canonical orbitals below (above) the Fermi level are
related to the hole components of hole-like (particle-like)
quasiparticle excitations.
However, in order to confirm it, we need quasiparticle states in the
full space, which are outside the scope of this paper.

Cross symbols plotted in Fig.~\ref{fig:wf_kappa} are 
$\kappa_i$ calculated at a farther radius ($r=20$ fm).
In this case, concerning the deepest and highest orbitals, 
$\kappa$ drops toward the minimum value at $\epsilon_i=\fermilevel$.
This may be ascribed to the fact that the canonical orbitals are constructed 
by mixing up the hole part of all the quasiparticle states
and thus the component having smallest $\kappa$ dominates in all the canonical
orbitals at large radii.
Probably, all the canonical orbitals have the same 
$\kappa$ 
($= \sqrt{2m(E_{\rm qp,min}-\epsilon_{\rm F})} /\hbar$
%
% ($\sim \sqrt{-2m\epsilon_{\rm F}} /\hbar$)
%
where $E_{\rm qp,min}$ is the smallest quasiparticle energy)
at sufficiently large radii.

%-------- Quasiparticle states (job285 + tjm:p/cbhfb/nx29hfb) ----------------

\subsection{Quasiparticle states \label{sec:cbqp}}

Let us discuss on the quasiparticle states defined by Eq.~(\ref{eq:hfb_matrix})
in the subspace spanned by $N_{\rm w}$ canonical orbitals.
These quasiparticles and the true quasiparticles defined in the full space are
quite different.  The former are only a set of tools for the minimization
in a small subspace, while the latter have physical information on
the excitation modes of the nucleus. 

The width $\Delta E_{\rm qp}^{(i)}$ (in the full space)  
of the excitation energy $E_{\rm qp}^{(i)}$ of the 
$i$th quasiparticle (in the subspace) is given by
\begin{equation}
\label{eq:deltaeqp} % -*- Eq : width of quasiparticle energy
\left( \Delta E_{\rm qp}^{(i)} \right)^2 =
\int \left( |\phi_i'(\vct{r})|^2 + |\varphi_i'(\vct{r})|^2 \right) \dvol
- \left(E_{\rm qp}^{(i)} \right)^2,
\end{equation}
\begin{displaymath}
\left( \begin{array}{c} \phi_i' \\ \varphi_i' \end{array} \right) =
\left( \begin{array}{cc} h - \fermilevel & \tilde{h} \\ 
\tilde{h} & - h + \fermilevel \end{array} \right)
\left( \begin{array}{c} \phi_i \\ \varphi_i \end{array} \right),
\end{displaymath}
where the operation of $h$ and $\tilde{h}$ should be evaluated
in the full space (i.e., in the mesh representation), 
not in the $N_{\rm w}$-dimensional subspace.

Quasiparticles can be characterized by 
the norms of the particle and hole components as well as by the
excitation energy. These norms are defined as
\begin{equation}
\label{eq:partholenorm} % -*- Eq : Norm of particle and hole components of qp
U_i^2 = \int |\phi_i(\vct{r})|^2 \dvol, \;\;\;
V_i^2 = \int |\varphi_i(\vct{r})|^2 \dvol,
\end{equation}
for the particle and hole components, respectively.
They are normalized as $U_i^2 + V_i^2 = 1$ according to Eq.~(\ref{eq:norm_qp}).
Quasiparticle states are called to be particle-like (hole-like)
if $U_i^2$ ($V_i^2$) $ > \frac{1}{2}$.

%-------------------------- F I G U R E -----------------------------------
%\begin{figure}[htb]%G:box
%\begin{center}\framebox{Figure \ref{fig:cbqpe}}\end{center}%G:box
\begin{figure}%G:psf
\includegraphics[angle=-90, width=8.6 cm]{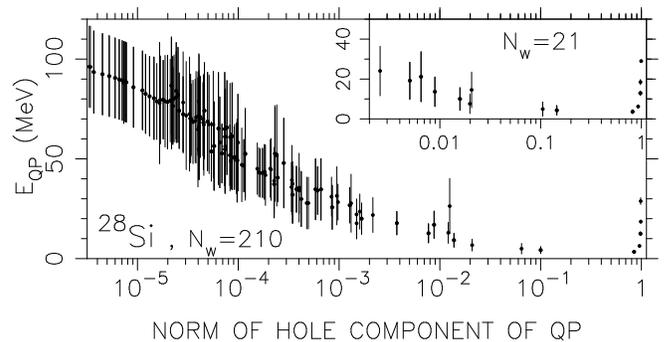}%G:psf
\caption{Excitation energies $E_{\rm qp}$ (dots) 
and their widths $\Delta E_{\rm qp} $(half of the length of vertical bars)
of quasiparticle states of $^{28}$Si 
plotted versus the norm of the hole component. 
Calculations are done using the HFB solution for $N_{\rm w}$=210 shown in
% Fig.~\ref{fig:canorb}.
Fig.~\figcanorb.
The top-right window shows the result for $N_{\rm w}$=21 for comparison.
}
\label{fig:cbqpe}
\end{figure}
%------------------- E N D   O F   F I G U R E ----------------------------

Figure \ref{fig:cbqpe} shows $E_{\rm qp}$ and $\Delta E_{\rm qp}$ for
$N_{\rm w}=210$ (with $v_{\rm p}=-837$ MeV fm$^3$).  One can see that
hole-like quasiparticles have very small widths.  Even the deepest hole
state (1s) has a relatively small width (1.7 MeV).  On the other hand,
the widths of particle-like quasiparticles amount to a few tens of MeV:
They are very different from the true quasiparticles defined in the full space.

For the sake of comparison, 
quasiparticle levels obtained in a smaller subspace 
($N_{\rm w}=21$, $v_{\rm p}=-1050$ MeV fm$^3$)
are also shown  in the top-right window.
By comparing the results of two calculations, one can see that
increasing the number of canonical orbitals leads to the addition of
high-excitation particle-like quasiparticles,
while it does not change very much the hole-like states and 
low-excitation particle-like states.
Thus, enlargement of the canonical-basis subspace is a very inefficient way to
decrease the energy widths of particle-like quasiparticle states.

The differences from true quasiparticles become more evident
by looking at the wave functions.
For $E_{\rm qp} > - \lambda$, the particle component $\phi$ should be
an oscillating function of $r$ \cite{DFT84}. 
However, $\phi$ of quasiparticles defined in the canonical-orbital
subspace has an exponentially damping tail.
It is only natural because they are expressed as linear combinations of
such functions.

It should be stressed again that the quasiparticles in the
canonical-orbital subspace are nevertheless a useful auxiliary tool 
to obtain the HFB ground state.
However, if one needs true quasiparticle excitation modes, 
the quasiparticles in the subspace are useless.
One has to calculate with some other method the true eigenstates of the
quasiparticle Hamiltonian.
A good news is that the Hamiltonian (being composed of $h$ and $\tilde{h}$) 
is already obtained and thus iterations for self-consistency are not necessary.
This will make the numerical calculations rather inexpensive.

%---------- Nw dependence of pairing gap (job270+278) ----------------------

\subsection{Choice of the value of $k_{\rm c}$}
% . . . . . . . . . . . . . . . . . . . . . . . . . . . . . . . . . . . . .

Our canonical-basis HFB method has several parameters 
to specify the pairing force, i.e.,
the over-all strength $v_{\rm p}$, 
the parameter for the momentum dependence $k_{\rm c}$, 
the number of canonical orbitals having non-zero occupation probabilities
$N_{\rm w}$,
and parameters for density dependences.
However, available experimental data
do not provide sufficient information on the pairing correlation
for precise determinations of so many parameters.
%
% The coherent length of nuclear pairing is much larger than
% the range of the interaction for nuclear pairing \cite{YB03}.
%
Therefore we tune only $v_{\rm p}$ precisely so that the resulting 
pairing gap has a desired value,
while assuming some reasonable values for the other parameters.
However, this does not guarantee that their values
may be chosen completely arbitrarily.
%
% One still has to examine the physical reasonableness
% One may also consider the convenience 
% for the employed method for solutions.
%
In this and next subsections we discuss on the choices of the values 
of $k_{\rm c}$ and $N_{\rm w}$, respectively.

Pairing-channel interactions were taken into account when
Skyrme forces SGII \cite{GS81} and SkP \cite{DFT84} were
determined.
Their values of $k_{\rm c}$ (=  $\sqrt{-t_0 (1-x_0)/t_1 (1-x_1)}$) are
2.5 fm$^{-1}$ and 4.3 fm$^{-1}$, respectively.
From this result, values from 2 to 5 fm$^{-1}$ seem equally reasonable.

From a pragmatic point of view to use the force
in the canonical-basis method, 
$k_{\rm c}$ must be smaller than 6 fm$^{-1}$
in order to avoid the point collapse (see Appendix~\ref{sec:pointcollapse}).
Values smaller than 1.5 fm$^{-1}$ are not preferable, either,
because such small $k_{\rm c}$ makes the solutions very difficult to obtain.

If $v_{\rm p}$ is adjusted for each value of $k_{\rm c}$
to reproduce the same value of the average pairing gap, 
which we define \cite{Taj99a} as ,
\begin{equation}  \label{eq:avegap}
\Delta = \sum_{i=1}^{N_{\rm w}} u_i v_i \Delta_i 
       \mbox{\Huge $/$} \sum_{i=1}^{N_{\rm w}} u_i v_i ,
\end{equation} 
resulting properties of the nucleus are almost independent of $k_{\rm c}$
except the pairing density, which becomes more diffuse when 
smaller $k_{\rm c}$ is used.
Unfortunately, pairing density is rather difficult to determine experimentally.

Incidentally,
the properties of each canonical orbital are rather sensitive to $k_{\rm c}$:
$v_i^2$ and $\langle  r_i^2 \rangle$ increase
while $\epsilon_i$ decreases with decreasing $k_{\rm c}$.
Only $\Delta_i$ stays roughly constant. 
Nevertheless, properties of the nucleus (except the pairing density
profile) are hardly changed,  as far as
the average pairing gap is kept unchanged.

%-------------------------- F I G U R E -----------------------------------
%\begin{figure}[htb]%G:box
%\begin{center}\framebox{Figure \ref{fig:nw_gap}}\end{center}%G:box
\begin{figure}%G:psf
\includegraphics[angle=-90, width=6.5 cm]{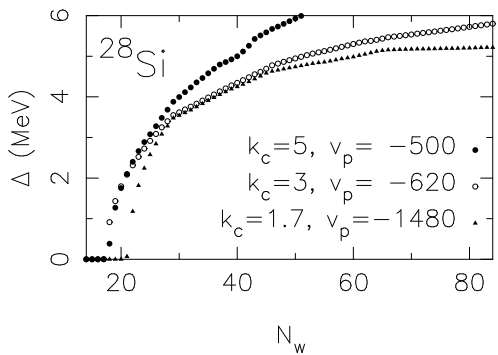}%G:psf
\caption{
Dependence of the average pairing gap $\Delta$ 
on the number of canonical orbitals $N_{\rm w}$
for three sets of pairing interaction parameters 
$k_{\rm c}$ (fm$^{-1}$) and $v_{\rm p}$ (MeV fm$^{3}$).
See text for explanations.
}
\label{fig:nw_gap}
\end{figure}
%------------------- E N D   O F   F I G U R E ----------------------------

From the allowed interval 1.5 fm$^{-1}$ $\le k_{\rm c} \le$ 6 fm$^{-1}$,
we have chosen 2 fm$^{-1}$ in this paper.
There is no physical reason for this choice.
However, smaller values of $k_{\rm c}$ lead to a favorable
situation that the dependence on $N_{\rm w}$ becomes weaker 
and the determination of $N_{\rm w}$ can be less precise.
Fig.~\ref{fig:nw_gap} shows the dependence of $\Delta$ of Eq.~(\ref{eq:avegap})
on $N_{\rm w}$.
Solid circles, open circles, and triangles are obtained with
$k_{\rm c}$= 5, 3, and $1.7$ fm$^{-1}$, respectively.
The values of $v_{\rm p}$ are given in the
legends of the figure in units of MeV fm$^3$.
Here, $v_{\rm p}$ is not changed as a function of $N_{\rm w}$ but
is determined for each $k_{\rm c}$ 
such that the resulting $\Delta$ is roughly the same at $N_{\rm w} \sim 28$.
One can see in the figure that, for $k_{\rm c}=1.7$ fm$^{-1}$,
$d \Delta / d N_{\rm w}$ is almost zero around $N_{\rm} \sim 80$.
The emergence of this plateau is explained by the vanishing 
of the pairing interaction at relative momentum $k_{\rm c}$.
For much larger $N_{\rm w}$,
$\Delta$ will increase again infinitely \cite{TOT94}. 

\subsection{Dependences on $N_{\rm w}$ \label{sec:uniquecanorb}}

\begin{figure}%G:psf
\includegraphics[angle=-90, width=6.25 cm]{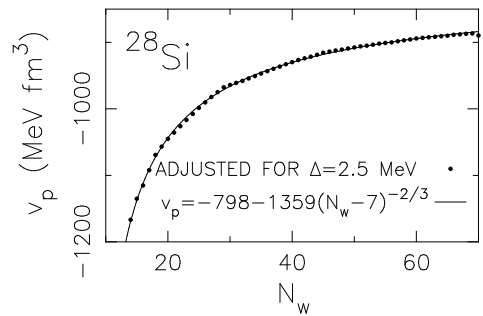}%G:psf
\caption{
Pairing force strength $v_{\rm p}$ adjusted to give
$\Delta = 2.5$ MeV as a function of
the number of canonical orbitals $N_{\rm w}$ for $^{28}$Si.
Dots denote the strengths obtained by the adjustment
while the curve represents a function fitted to the dots.
}
\label{fig:nw_pfs}
\end{figure}

We next study the dependence on the number of canonical orbitals
$N_{\rm w}$. Here we adjust the strength $v_{\rm p}$
for each value of $N_{\rm w}$ to keep $\Delta$ constant.
The used values of $v_{\rm p}$ are shown in Fig.~\ref{fig:nw_pfs}.

%-------------------------- F I G U R E -----------------------------------
%\begin{figure}[htb]%G:box
%\begin{center}\framebox{Figure \ref{fig:nw_pfs}}\end{center}%G:box
%------------------- E N D   O F   F I G U R E ----------------------------

%-------------------------- F I G U R E -----------------------------------
%\begin{figure}[htb]%G:box
%\begin{center}\framebox{Figure \ref{fig:nw_etot}}\end{center}%G:box
\begin{figure}%G:psf
\includegraphics[angle=-90, width=7.0 cm]{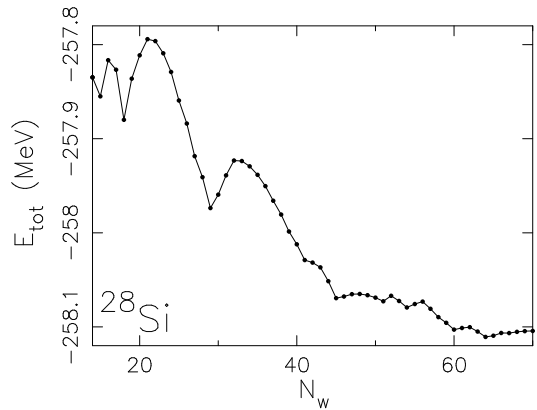}%G:psf
\caption{
Dependence of the total energy $E_{\rm tot}$
on the number of canonical orbitals $N_{\rm w}$ for $^{28}$Si.
See text for explanations.
}
\label{fig:nw_etot}
\end{figure}
%------------------- E N D   O F   F I G U R E ----------------------------

In Fig.~\ref{fig:nw_etot} 
we show the total energy of $^{28}$Si as a function of $N_{\rm w}$.
It is obtained as follows:
Starting from harmonic oscillator eigenstates, we obtain an HFB solution
for $N_{\rm w}=70$. Then, we remove the canonical orbital with the
smallest $v_i^2$ and again obtain the HFB solution for $N_{\rm w}=69$ by the
gradient method.
By repeating this procedure, we obtain solutions for $N_{\rm w} \ge 14$.
This figure shows that the magnitude of the influence of $N_{\rm w}$
on the total energy is 300 keV or less
when $v_{\rm p}$ is determined for the average pairing gap.
Other bulk properties of the nucleus are also roughly 
independent of $N_{\rm w}$:
Deformation parameters $\beta$ and $\gamma$ are
within $0.374 \pm 0.01$ and  $\pm 0.3^{\circ}$, respectively.
R.m.s mass radius is $3.201 \pm 0.004$ fm.

%-------------------------- F I G U R E -----------------------------------
%\begin{figure}[htb]%G:box
%\begin{center}\framebox{Figure \ref{fig:nw_spl}}\end{center}%G:box
\begin{figure}%G:psf
\includegraphics[width=6.9 cm]{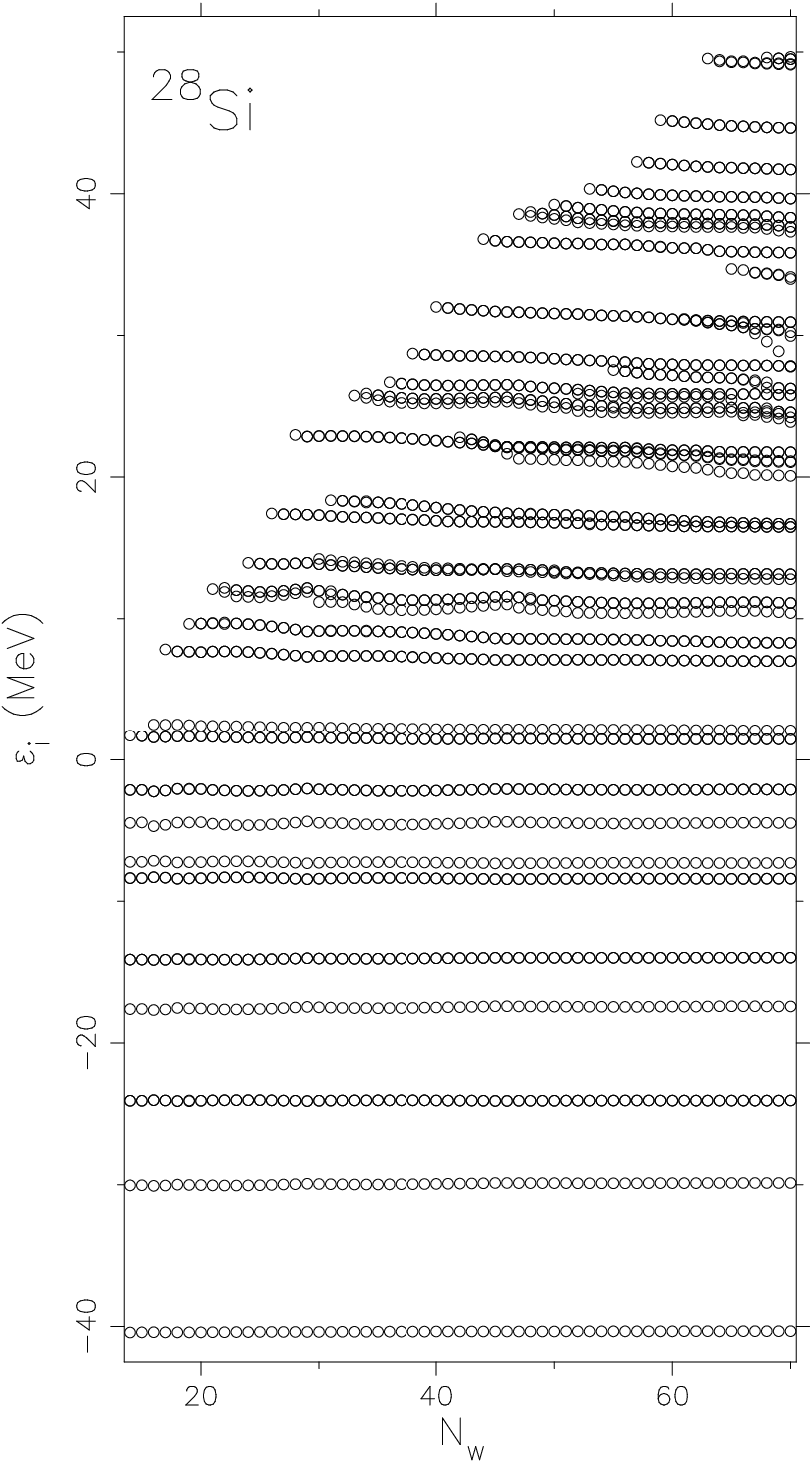}%G:psf
\caption{
Dependence of the mean-field energies $\epsilon_i$ of the canonical orbitals
on the number of the orbitals $N_{\rm w}$ for $^{28}$Si.
See text for explanations.
}
\label{fig:nw_spl}
\end{figure}
%------------------- E N D   O F   F I G U R E ----------------------------

Such insensitivity can also be seen in the properties of individual
canonical orbitals.
In Fig.~\ref{fig:nw_spl},
mean-field energies  $\epsilon_i$ of canonical orbitals
are plotted versus $N_{\rm w}$.
One can see that $\epsilon_i$ below 5 MeV stays almost constant.
Higher levels are not so constant but only slightly decreasing.
We have also confirmed that other quantities also
remain roughly constant.

Therefore, unlike $k_{\rm c}$, the effects of changing $N_{\rm w}$ is
almost canceled by the adjustment of $v_{\rm p}$.

%---------- Mass number dependence (job282) ----------------------------------

\subsection{Application to heavy nuclei \label{sec:mass_dep}}

Finally we demonstrate the applicability of
the canonical-basis HFB method to nuclei heavier
than $^{28}$Si.
We have encountered no special difficulties at least up to $A=252$
except the increase in the computation time per gradient-method step.

\begin{figure}%G:psf
\includegraphics[angle=-90, width=8.6 cm]{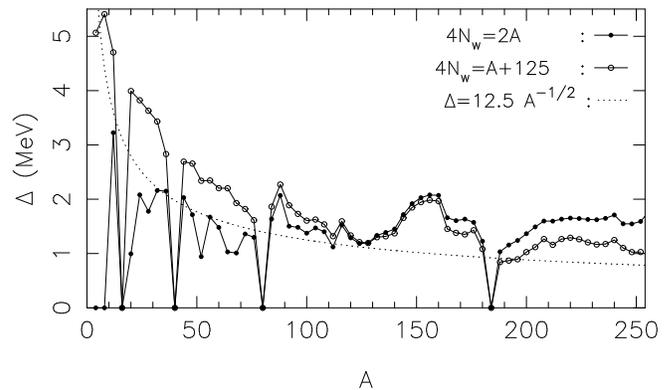}%G:psf
\caption{
Dependence of the average pairing gap $\Delta$ on the mass number $A$.
The strength of the pairing interaction is fixed at 
$v_{\rm p}=-1000$ MeV fm${^3}$ while $N_{\rm w}$ is changed in 
two ways as shown in the legend.
}
\label{fig:mass_gap}
\end{figure}

For example,
Fig.~\ref{fig:mass_gap} shows the dependence of the
average pairing gap $\Delta$ on the mass number $A$.
In the calculations, 
the normalization box is gradually expanded with increasing $A$:
The edge of the box is 
%
% the larger of $6 A ^{1/3}$ fm and 16 fm, i.e.,
%
at least 2.5 times as large as the liquid-drop-model diameter.
We fix $v_{\rm p}$ at  $-1000$ MeV fm${^3}$.
Instead, we change $N_{\rm w}$ versus $A$.

When we assume a relation $4N_{\rm w} = 2A$,
we cannot reproduce the empirical trend that $\Delta$
is a decreasing function of $A$.
We can improve the result by using different relations, e.g., 
$4 N_{\rm w} = A+125$.
However, the goal, a universal pairing force, should be more than
an empirical formula expressing $N_{\rm w}$ as a function of $A$.
It will be one of our future challenges.

%-------------------------- F I G U R E -----------------------------------
%\begin{figure}[htb]%G:box
%\begin{center}\framebox{Figure \ref{fig:mass_gap}}\end{center}%G:box
%------------------- E N D   O F   F I G U R E ----------------------------

%====================== C O N C L U S I O N S ================================

\section{Conclusions \label{conclusions}}

In this paper we have presented a method to obtain solutions of the
HFB equation expressed in the canonical form, i.e. in terms of BCS-type
wave functions, without using quasiparticle states.
The method is fit to three-dimensional coordinate space representations
and is advantageous to describe simultaneously arbitrary deformations,
long and short density tails, 
and pairing correlations involving states in the continuum of the
HF Hamiltonian.

We have improved the speed of the convergence to HFB solutions
by a few orders of magnitude.
For this purpose, we have modified the gradient method and
derived the appropriate form of the Lagrange
multiplier functionals for the orthogonality between canonical orbitals.
The two-basis method has also been adapted by replacing the HF basis
with the canonical basis to further speedup the convergence.

For delta-function pairing interactions, the wave functions of
canonical orbitals have been shown to shrink infinitely into a spatial
point once their occupation probabilities become less than some critical
value.  This problem is peculiar to the canonical-basis HFB method.  
To avoid it, a repulsive momentum dependent term has been added to the
pairing interaction. 

As a byproduct of the introduction of this term, the nature of
high-lying canonical orbitals has been greatly clarified
as approximate eigenstates of the pairing Hamiltonian.  
We have demonstrated that the level density and the spatial extent of
high-lying canonical orbitals can be computed by applying the
Thomas-Fermi approximation to the pairing Hamiltonian.

The obtained canonical orbitals have been examined in many aspects.
One of the results is that canonical orbitals closer to the Fermi level
have wave functions with longer tails. 
At very large radius, however, all the orbitals seem to change the slope
of their tails to that of the orbital closest to the Fermi level.
Another result is that quasiparticle states cannot be efficiently
expanded in the canonical basis. 
Therefore the canonical orbitals perfectly describing the ground state 
are not sufficient to treat excitations.

The pairing force strength has been adjusted to reproduce the average
pairing gap as a function of the number of canonical orbitals,
which is used instead of an energy cut-off.
Properties of the nucleus have been found to be
almost independent of the number of canonical orbitals and
rather insensitive to the relative strength of the momentum dependent to
independent terms of the pairing force.

Although most of our discussions refer to calculations for $^{28}$Si,
we have also checked the applicability of the method 
to heavy nuclei up to $A$=252.
Checks have also been done concerning the precision of the mesh
representation in connection with this method.

% - - - - - - - - - - - - - - - - - - - - - - - - - - - - - - - - - - -

We believe that the canonical-basis HFB method has the potential to
become the standard approach to treat neutron-rich nuclei in HFB.  
As a next step, we are going to extend the formalism and the computer
program concerning the treatment of $N \not= Z$ nuclei and the inclusion
of the spin-orbit and the Coulomb interactions.

%====================== A P P E N D I X ===================================
\appendix  % Use \appendix* if there is only one appendix.

%--------------------------- Appendix A --------------------------------
\section{Precision of the mesh representation \label{sec:meshprec}}

In this Appendix, we examine the precision of the Cartesian mesh
representation when it is applied to the canonical-basis HFB method.
Discussions are given concerning the dependencies on i) the mesh spacing,
ii) the approximation formulae for the derivatives,
and iii) the size of the normalization box.
The pairing force strength used is $v_{\rm p} = -1032$ MeV fm$^3$ 
with $N_{\rm w}=21$.
The state is the prolate ground state of $^{28}$Si.

%-------- Mesh spacing dependence (job271+279+283) -----------------------

i) We discuss the dependence on the mesh spacing $a$.
For this purpose, we calculate the HF and HFB solutions
for various value of $a$ while keeping the size of the box constant.
For the discrete approximation of the first and second derivatives, 
we use the 17-point formula \cite{Taj01a}.
According to the calculation using the finest mesh ($a=0.3$ fm),
$E_{\rm tot}=-257.723$ MeV, $\Delta=2.424$ MeV, and $\beta=0.3766$.

The mesh is changed in the following manner:
We use $\nmesh$ mesh points in $x$, $y$, and $z$-directions.
Then the length of the edge of the normalization box is $L$ = $(\nmesh + 1)a$.
We fix $L$ at 24 fm so that increasing $\nmesh$ from 15 to 79 corresponds to
decreasing $a$ from 1.5 fm to 0.3 fm.
The error is defined as the deviation from the result with $a$=0.3 fm.
The center of mass of the nucleus is placed at the center of the box. 

%-------------------------- F I G U R E -----------------------------------
%\begin{figure}[htb]%G:box
%\begin{center}\framebox{Figure \ref{fig:mesh_dep}}\end{center}%G:box
\begin{figure}%G:psf
\includegraphics[angle=-90, width=7.0 cm]{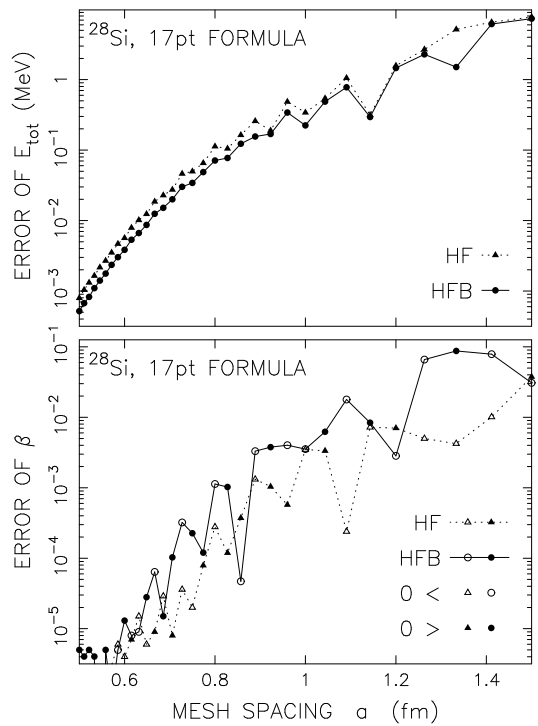}%G:psf
\caption{
Errors of the total energy $E_{\rm tot}$ (top portion)
and the quadrupole deformation parameter $\beta$ (bottom portion)
due to the finite mesh spacing $a$ for the prolate solution of $^{28}$Si.
The result of HF (HFB) calculations are designated by
triangles (circles) connected with dot (solid) lines.
In the bottom portion, triangles and circles are open (solid) when the 
error is positive (negative).
}
\label{fig:mesh_dep}
\end{figure}
%------------------- E N D   O F   F I G U R E ----------------------------

The top portion of Fig.~\ref{fig:mesh_dep} shows the error of the total
energy $E_{\rm tot}$ as a function of the mesh spacing $a$.  
The error of the corresponding HF calculation is also shown for comparison.
Before obtaining this result,
we suspected that HFB solutions suffer from larger errors
than HF solutions because the former involve states above the Fermi
level while the latter depend on only levels below it.
Higher levels contain more amount of high momentum components which 
are less accurately treated with a finite mesh spacing.
However, this figure shows that the errors are of the same order.
The principal reason may be that 
errors from high-lying levels are reduced by factors $v_i^2$ or $u_i v_i$.

The bottom portion of Fig.~\ref{fig:mesh_dep} shows the error of the 
quadrupole deformation parameter $\beta$.
Unlike $E_{\rm tot}$, the error of $\beta$ takes both positive and
negative values.
For this quantity, the HFB errors are several times as large as 
the HF errors on the average.
This may be related to the strong influence of the pairing correlation
on deformations.

%-------- Approximation formula dependence (job276) -----------------------

ii) We compare some approximation formulae given in Ref.~\cite{Taj01a}
for the evaluation of the derivatives.
Note that the Fourier transformation method 
(or the Lagrange-mesh method \cite{BH86} applied to uniform-spacing meshes)
assumes a different boundary condition
from the other formulae. It assumes the periodic boundary condition 
and the edge of the box is also slightly shortened to $\nmesh a$.
The error is defined as the deviation from the most precise result,
which is obtained with the 17-point formula and $a$=0.3 fm.  We have not
chosen the Fourier method for this purpose because of a large error due
to numerical cancellations for more than several tens of mesh points.

%-------------------------- F I G U R E -----------------------------------
%\begin{figure}[htb]%G:box
%\begin{center}\framebox{Figure \ref{fig:fml_dep}}\end{center}%G:box
\begin{figure}%G:psf
\includegraphics[angle=-90, width=7.5 cm]{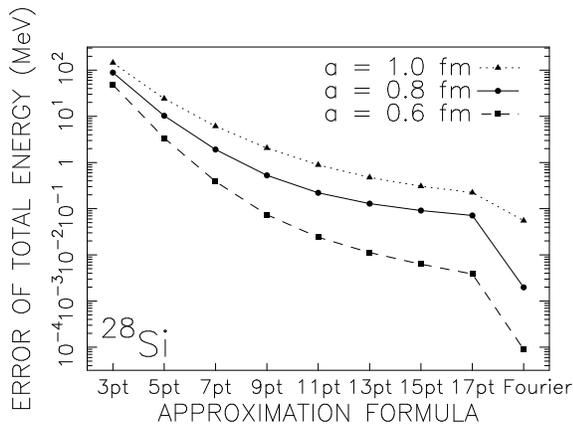}%G:psf 
\caption{
Dependence of the error of the HFB total energy $E_{\rm tot}$ of $^{28}$Si
on the discrete approximation formulae used to evaluate the
derivatives. Three curves correspond to different values of the mesh
spacing $a$.
} 
\label{fig:fml_dep}
\end{figure}
%------------------- E N D   O F   F I G U R E ----------------------------

In Fig.~\ref{fig:fml_dep}, the error of the total energy is shown 
for each approximation formula and for three values of the mesh spacing $a$.
This figure tells, e.g., if one uses the 17-point formula with $a=0.8$ fm
for $^{28}$Si, the error of the total energy is 70 keV.

One can observe two trends: 
First, for all the three mesh spacings, the error decreases as the
formula involves more number of mesh points 
(Fourier transformation method uses all the mesh points in the box).
Second, for every formula, 
the error decreases when the mesh spacing is decreased.
Therefore, one can decrease the error of the total energy either by
employing a more precise formula or by diminishing the mesh spacing.

On the other hand, the other quantities cannot always be improved
significantly by using more precise formulae.
For example, with $a$=1.0 fm,
the error of the pairing gap $\Delta$ does not
continue to decrease but seems to almost saturate around 10 keV
for more than 9-point formulae (including the Fourier transformation).
We have also found a similar saturation phenomenon in the error of the 
deformation parameter $\beta$, which is affected strongly by the pairing gap.
The reason of this saturation is probably that the pairing gap is more
sensitive than the total energy to high-lying states having large
momentum and thus the lack of momentum larger than $\pi / a$ in the mesh
space becomes the main source of the error, while the change of the
formulae can only improve the treatment of momentum components smaller 
that $\pi / a$.

%------------- Box size dependence (job272+280) ---------------------------

iii) We examine the errors due to the finite size of the normalization box.
We consider $^{28}$Si, which has $\fermilevel = -11.6$ MeV.
Naturally the errors depend on the Fermi level.
In this paper, however, we can treat only $N=Z$ nuclei, all of which have
roughly the same Fermi level.

We control the size of the box $L$ in terms of $\nmesh$ 
while fixing $a$ at 0.8 fm.
We define the errors as the deviation from the result obtained with 
large $L$ ($\sim 60$ fm).
Here, the comparison should be made within the same parity of 
$\nmesh$ \cite{IH03}.

%-------------------------- F I G U R E -----------------------------------
%\begin{figure}[htb]%G:box
%\begin{center}\framebox{Figure \ref{fig:box_dep}}\end{center}%G:box
\begin{figure}%G:psf
\includegraphics[angle=-90, width=8.6 cm]{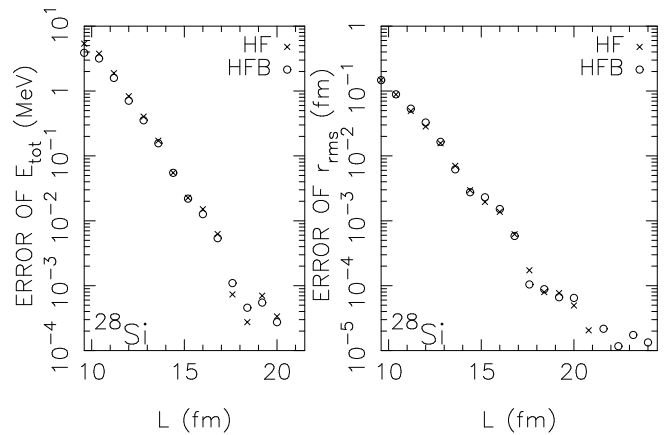}%G:psf
\caption{
Errors of the total energy $E_{\rm tot}$ (on the left-hand side)
and the r.m.s.\ mass radius $r_{\rm rms}$ (on the right-hand side) due to
the finite box size $L$.
}
\label{fig:box_dep}
\end{figure}
%------------------- E N D   O F   F I G U R E ----------------------------

On the left- and right-hand sides of Fig.~\ref{fig:box_dep}, 
we show the errors of the total energy $E_{\rm tot}$
and the r.m.s.\ radius, respectively.
We have used $L=24$ fm in most of the calculations shown in this paper.
These figures confirm that this value of $L$ leads to sufficiently 
precise results for both quantities.
By comparing the errors of HF (crosses) and HFB (circles) solutions,
one can see that the effects of the finite box size are of the same order 
between the two methods.
This is because the density tail is determined by the
Fermi level, which is roughly the same whether there exist pairing
correlations or not.

%------------------------- Appendix B --------------------------------------

\section{The mechanism of the collapse of a high-lying canonical orbital
into a spatial point and its remedies \label{sec:pointcollapse}}

In this appendix we study the mechanism of ``point collapse'', 
a phenomenon that the wave function of a canonical orbital shrinks
infinitely into a point in the coordinate space once
its occupation probability becomes less than some critical value.
It is a problem peculiar to the canonical-basis HFB method.
For this purpose, we consider only one canonical orbital
explicitly while representing all the others in terms of
uniform densities, $\rho_0$, $\tilde{\rho}_0$, $\tau_0$, and $\tilde{\tau}_0$.
The explicitly considered orbital is assumed to have a wave function
whose spatial part is a Gaussian wave packet
parameterized with a size parameter $w$ as,
\begin{equation} \label{eq:wavepacket}
\psi_w(\bm{r}) = \left(\frac{2}{\pi}\right)^{3/4} w^{-3/2} e^{-(r/w)^2}.
\end{equation}
The r.m.s.\ values of the coordinates and the wave numbers 
of this wave packet are,
respectively, 
$\sqrt{\langle x^2 \rangle}$ = 
$\sqrt{\langle y^2 \rangle}$ = 
$\sqrt{\langle z^2 \rangle}$ = 
$\frac{w}{2}$ and
$\sqrt{\langle k_x^2 \rangle}$ = 
$\sqrt{\langle k_y^2 \rangle}$ = 
$\sqrt{\langle k_z^2 \rangle}$ = 
$\frac{1}{w}$.
One can express the densities of the total system as,
\begin{eqnarray}
\rho & = \rho_0 + \rho_1, \;\;\;  &
\rho_1 = 4 v^2 \left| \psi_w (\bm{r})\right|^2, \\
\tau & = \tau_0 + \tau_1, \;\;\;  &
\tau_1 = 4 v^2 \left| \nabla \psi_w (\bm{r})\right|^2, \\
\tilde{\rho} & = \tilde{\rho_0} + \tilde{\rho_1}, \;\;\;  &
\tilde{\rho_1} = 4 uv \left| \psi_w (\bm{r})\right|^2, \\
\tilde{\tau} & = \tilde{\tau_0} + \tilde{\tau_1}, \;\;\;  &
\tilde{\tau_1} = 4 uv \left| \nabla \psi_w (\bm{r})\right|^2 ,
\end{eqnarray}
where $v^2$ is the occupation probability of this orbital and $u=\sqrt{1-v^2}$.

The change in the total energy due to the presence of the 
explicitly considered orbital is given by
\begin{equation}
\Delta E_{\rm tot} = 4 \pi \int_0^{\infty} 
\left[ {\cal H}(r) - {\cal H}_0 (r) \right] r^2 dr,
\end{equation}
where ${\cal H}$ has been defined by Eq.~(\ref{eq:ham_dens}), while
${\cal H}_0$ by replacing $\rho$, $\tau$, $\tilde{\rho}$, and $\tilde{\tau}$ 
in the equation with
$\rho_0$, $\tau_0$, $\tilde{\rho}_0$, and $\tilde{\tau}_0$, respectively.
For $k_{\rm c}$ = $\tilde{\rho}_{\rm c}$ = $\infty$,
one can obtain the following expression for $\Delta E_{\rm tot}$
($\alpha = 1$ is assumed for the density dependence of the Skyrme force):
\begin{eqnarray}
\Delta E_{\rm tot} & = & 
  C_1 uv 
+ C_2 v^2 
+ C_3 \frac{v^2}{w^2}
+ C_4 \frac{u^2 v^2}{w^3}
+ C_5 \frac{u v^3}{w^3} \nonumber \\
& + & C_6 \frac{v^4}{w^3}
+ C_7 \frac{v^4}{w^5}
+ C_8 \frac{u^2 v^4}{w^6}
+ C_9 \frac{v^6}{w^6}, \label{eq:DEtot}
\end{eqnarray}
where
\begin{eqnarray}
C_1 & = & v_{\rm p} \left( 1-\frac{\rho_0}{\rho_{\rm c}} \right) 
          \tilde{\rho}_0 ,\nonumber \\
C_2 & = & \nonumber 4 \left(f_1 \tau_0 + \frac{3}{4}t_0 \rho_0 +\frac{3}{16}
          t_3 \rho_0^2 - \frac{v_{\rm p}}{8\rho_{\rm c}}\tilde{\rho_0}^2 
          \right) ,\nonumber\\
C_3 & = & 12\left( \frac{\hbar^2}{2m}+f_1 \rho_0 \right) , \nonumber \\
C_4 & = & \frac{2}{\left( \sqrt{\pi} \right)^3} v_{\rm p} 
          \left( 1-\frac{\rho_0}{\rho_{\rm c}} \right) , \nonumber \\
C_5 & = & -\frac{4}{\left( \sqrt{\pi} \right)^3} v_{\rm p} 
           \frac{\tilde{\rho}_0}{\rho_{\rm c}} , \nonumber \\
C_6 & = & \frac{3}{\left( \sqrt{\pi} \right)^3}
          \left(2t_0+t_3\rho_0\right) , \nonumber \\
C_7 & = & \frac{24}{\left( \sqrt{\pi} \right)^3}
          \left(f_1-4f_2\right) , \nonumber \\
C_8 & = & -\frac{64}{\left( \sqrt{3} \pi \right)^3}
           \frac{v_{\rm p}}{\rho_{\rm c}} , \nonumber \\
C_9 & = & \frac{32}{\left( \sqrt{3} \pi \right)^3} t_3 . \nonumber
\end{eqnarray}
Among the nine coefficients, 
$C_1, C_2, C_4,$ and $ C_6$ are negative (at low density places) while
the others are positive.

%----- dip of the potential --------------------------------------------

%-------------------------- F I G U R E -----------------------------------
%\begin{figure}[htb]%G:box
%\begin{center}\framebox{Figure \ref{fig:shrink}}\end{center}%G:box
\begin{figure}%G:psf
\includegraphics[angle=-90, width=7.5 cm]{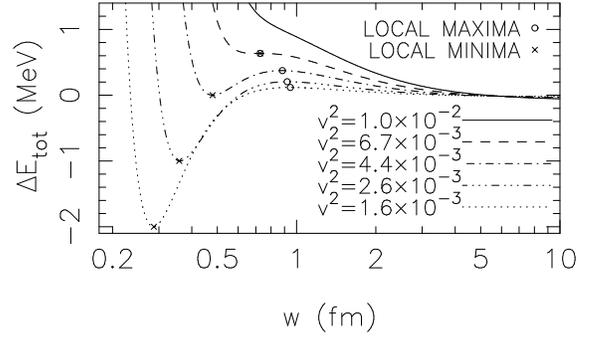}%G:psf
\caption{
Change in the total energy $\Delta E_{\rm tot}$ due to the addition of
a canonical orbital with occupation probability $v^2$ 
whose wave function is a Gaussian wave packet parameterized with a width 
parameter $w$.
$\Delta E_{\rm tot}$ is plotted versus $w$ for several values of $v^2$.
Local maxima and minima of the curves are designated with circles and crosses,
respectively.
}
\label{fig:shrink}
\end{figure}
%------------------- E N D   O F   F I G U R E ----------------------------

Fig.~\ref{fig:shrink} shows $\Delta E_{\rm tot}$ 
as a function of $w$ for several values
of $v^2$.
In the calculations, we employ
a pairing force having $v_{\rm p} = -450$ MeV fm$^{3}$,
$\rho_{\rm c}=0.32$ fm$^{-3}$, and $\tilde{\rho}_{\rm c}=k_{\rm c}=\infty$.
This strength $v_{\rm p}$ would give $\Delta=2$ MeV 
for $A=28$ system with $N_{\rm w}=21$ if point collapses would not occur.
% [ See job277 for the extrapolation to k_c=infinity ]
%
The effects of the momentum and the pairing density dependent forces are
discussed later.
The mean-field force is SIII without the spin-orbit term.
$A= \infty$ is used in Eq.~(\ref{eq:nucleonmass}) for the nucleon
reduced mass.
For the background densities, we employ $\rho_0 = \tilde{\rho}_0 = 1
\times 10^{-3}$ fm$^{-3}$ and $\tau_0 = \tilde{\tau}_0 = 3.6 \times
10^{-5}$ fm$^{-5}$, which are typical values at peripheral regions
where point collapses occur in actual HFB calculations for finite nuclei.
(A Fermi gas relation $\tau = \frac{3}{5}(\frac{3 \pi^2}{2})^{2/3} \rho^{5/3}$ 
is assumed to determine $\tau$ from $\rho$.)

One can see from Fig.~\ref{fig:shrink} that
the total energy becomes minimum at the bottom of a dip
for small values of $v^2$.
The dip emerges at $w = 0.7$ fm when $v^2$= $6.7 \times 10^{-3}$
and,  as $v^2$ is decreased, its depth increases while the location 
of the bottom moves to smaller values of $w$.

This minimum in the dip corresponds to
HFB ``solutions'' in finite nuclei reported in Refs.~\cite{Taj99a,Taj00a},
in which the wave function of a high-lying orbital has shrunken 
into a mesh point.
HFB states having physically reasonable wave functions turn themselves into
such strange states in the following abrupt way:
In the course of gradient-method evolutions, 
suddenly after an almost stationary situation has continued,
$\epsilon$ of a high-lying orbital soars up to a few hundred MeV
and simultaneously the total energy decreases by a few MeV.

The dip is caused by the zero-rangeness of the pairing force
and thus does not correspond to physically meaningful states.
Physical solutions should be found at $w>6$ fm, 
where $\Delta E_{\rm tot}$ is slightly negative.
Therefore one needs to eliminate the effects of such dips
in order to obtain well-behaving HFB solutions
in the canonical-basis representation.

It is worth pointing out here that a wave packet with $w$=0.3 fm
can be expressed fairly precisely on a mesh with spacing $a$=0.8 fm.
It is because the maximum wave number expressible with the mesh
is $\frac{\pi}{a}$ and thus
the minimum size of the wave packet of Eq.~(\ref{eq:wavepacket}) can be
estimated roughly by a relation
$\sqrt{\langle k_x^2 \rangle}$ = $\frac{1}{w}$ $\sim$ $\frac{\pi}{a}$,
which leads to $w \sim \frac{a}{\pi} = 0.26$ fm for $a$=0.8 fm.
Therefore collapses to wave packets as small as $w=0.3$ fm 
are a practical problem even when $a$=0.8 fm.

%----- local minimum -------------------------------------------------

Before discussing how to remove the dip,
let us clarify in an analytical way the mechanism which gives rise to the dip.
We consider the limit of $v^2, w \rightarrow 0$,
in which the nature of the dip becomes most manifest.
Therefore, the following considerations are for the untruncated Hilbert space, 
not for mesh representations.

In order to study this limit,
we assume a scaling behavior $v = \alphaip w^{\betaip}$
where $w$ is the location of the bottom of the dip.
$\alphaip$ and $\betaip$ are positive constants to be determined.
One can easily confirm that,
for $\betaip > \frac{3}{2}$,
all the nine terms in the right-hand side of Eq.~(\ref{eq:DEtot})
have positive powers of $w$ and thus
converge to zero as $w \rightarrow 0$.
Consequently, it must be satisfied that $\betaip \le \frac{3}{2}$
for a point-collapse solution to exist.
For $0 < \betaip < \frac{3}{2}$, 
$\Delta E_{\rm tot} = C_8 \alphaip^4 \left( \frac{1}{w} \right)^{6-4\betaip}$
+ (lower order terms in $\frac{1}{w}$)
and thus the energy diverges to $+\infty$ as $w \rightarrow 0$.
The only possibility for a point-collapse solution to exist
is in the case of $\betaip = \frac{3}{2}$, for which
$\Delta E_{\rm tot} = C_4 \alphaip^2 + C_8 \alphaip^4$ 
+ (terms with positive powers in $w$).
From a condition $\partial \Delta E_{\rm tot}/\partial \alphaip = 0$,
one immediately finds that $\Delta E_{\rm tot}$ becomes minimum
at $\alphaip^2 = -C_4 / 2 C_8$.
The minimum value in the limit $w \rightarrow 0$ becomes
\begin{equation} \label{eq:localmin}
\Delta E_{\rm tot} = -\frac{(C_4)^2}{4 C_8}
= \frac{\left(\sqrt{3}\right)^3}{64}v_{\rm p}\rho_{\rm c}\left(
1-\frac{\rho}{\rho_{\rm c}}\right)^2.
\end{equation}
Its numerical value is $-11.7$ MeV for the parameters used.
The essential points clarified here are 
i) the minimum of the total energy is realized
in the limit $v, w \rightarrow 0$ with $v \propto w^{3/2}$
and ii) the limiting value of the minimum energy is finite.

%----- local maximum -------------------------------------------------

One can also derive analytically the location $w$ of the barrier top
which divides the physical and unphysical regions.
From the condition,
\begin{equation}
\frac{\partial \Delta E_{\rm tot}}{\partial w} = -2C_3 \frac{v^2}{w^3}
-3C_4 \frac{u^2 v^2}{w^4} + {\cal O} (v^3) = 0,
\end{equation}
one can simply obtain the location of the local maximum as,
\begin{equation} \label{eq:localmax}
w = - \frac{3 C_4}{2 C_3}
  = - \frac{v_{\rm p} \left(1-\frac{\rho_0}{\rho_{\rm c}}\right)}
{ 4 \left(\sqrt{\pi}\right)^3 \left(\frac{\hbar^2}{2m} + f_1 \rho_0 \right)},
\end{equation}
by neglecting terms of ${\cal O} (v^3)$.
Note that it is independent of $v$ (if $v$ is sufficiently small). 
With the parameters used, $w = 1.0$ fm.
Indeed in Fig.~\ref{fig:shrink}, 
the location of the top of the barrier (designated with open circles)
seems to converge to $\sim 1$ fm as $v^2$ is decreased.
The height of the maximum is $C_1 v + {\cal O}(v^2)$, 
which is ${\cal O}(v)$ and
converges to zero as $v \rightarrow 0$.
Therefore, for sufficiently small values of $v$, 
HFB solutions easily fall into the unphysical dip
since the mesh spacing is much smaller than $1.0 \pi$ fm.

%----- effects of the kc term ----------------------------------------

Now, we examine two types of pairing interactions which can remove
the unphysical dip.

The first type of the interaction is 
the momentum dependent term of Eq.~(\ref{eq:pair_int}).
With $k_{\rm c} < \infty $, but with $\tilde{\rho}_{\rm c} = \infty$,
$\Delta E_{\rm tot}$ becomes to have additional terms as,
\begin{equation} \label{eq:DEtotkc}
\Delta E_{\rm tot}' = \Delta E_{\rm tot}
                    + C_{10} uv  
                    + C_{11} \frac{uv}{w^2}
                    + C_{12} \frac{u^2 v^2}{w^5} ,
\end{equation}
where $\Delta E_{\rm tot}$ is given by Eq.~(\ref{eq:DEtot}) and,
\begin{eqnarray}
C_{10} & = & - \frac{v_{\rm p}}{2 k_{\rm c}^2} \tilde{\tau}_0 , \nonumber \\
C_{11} & = & - \frac{v_{\rm p}}{2 k_{\rm c}^2} 3 \tilde{\rho}_0 , \nonumber \\
C_{12} & = & - \frac{v_{\rm p}}{2 k_{\rm c}^2}
               \frac{30}{\left(\sqrt{\pi}\right)^3} . \nonumber
\end{eqnarray}
Since $v_{\rm p} < 0$, $C_{10}$, $C_{11}$, and $C_{12}$ are positive.
The dominant term in the limit $w \rightarrow 0$ is
the one having coefficient $C_{11}$ ($C_{12}$) for $\betaip > (<) 3$,
which behaves as $\alphaip w^{\betaip-2}$ ($\alphaip^2 w^{2\betaip-5}$).
These two terms overwhelm the negative terms in  $\Delta E_{\rm tot}$. 
Consequently wave functions cannot shrink to a point.

%-------------------------- F I G U R E -----------------------------------
%\begin{figure}[htb]%G:box
%\begin{center}\framebox{Figure \ref{fig:shrinkkc}}\end{center}%G:box
\begin{figure}%G:psf
\includegraphics[angle=-90, width=8.6 cm]{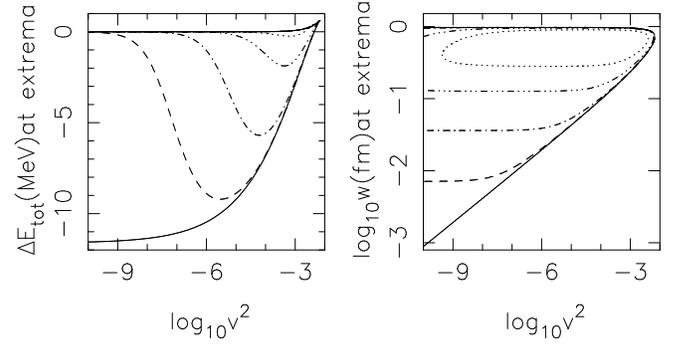}%G:psf
\caption{
Effects of changing $k_{\rm c}$ on the
total energy $\Delta E_{\rm tot}$ (left-hand portion)
and on the width parameter $w$ of the added orbital 
at local minimum and maximum of $\Delta E_{\rm tot}$.
The dot, triple-dot-dash, dot-dash, dash, and solid curves are obtained with
$k_{\rm c} = 15, 30, 100, 500$, and $\infty$ (fm$^{-1}$), respectively.
The other parameters of the model are the same as those used for
Fig.~\figshrink.
}
\label{fig:shrinkkc}
\end{figure}
%------------------- E N D   O F   F I G U R E ----------------------------

However, a dip can still emerge with finite $v$.
Figure~\ref{fig:shrinkkc} shows how the local minimum
(the bottom of the dip) and the local maximum (the top of the 
barrier) of $\Delta E_{\rm tot}$ change versus $v^2$.
The left-hand portion shows the depth and the height ($\Delta E_{\rm tot}$), 
while the right-hand portion shows the locations ($w$).

The solid curves are obtained with $k_{\rm c}=\infty$.
The upper solid curve stands for the height of the local maximum,
which converges to zero as $v^2 \rightarrow 0$,
as already discussed analytically.
(This curve is almost unchanged
 when $k_{\rm c}$ is decreased to finite values.)
The lower solid curve corresponds to the depth of the local minimum, which
converges to $-11.7$ MeV, in accord with Eq.~(\ref{eq:localmin}).

With finite values of $k_{\rm c}$, the minimum becomes
shallower as $v^2$ is decreased and it disappears at some small value of
$v^2$.  This disappearance is clearly seen in the dot curve 
(corresponding to $k_{\rm c}$=15 fm$^{-1}$)
in the right-hand portion of the figure.
A numerical search has shown that the dip exists only for
$k_{\rm c} > 10$ fm$^{-1}$.

The result with this simplified model roughly agrees
with realistic HFB calculations:
We have observed that point collapses can occur for $k_{\rm c} > 6 $ fm$^{-1}$
in calculations like those shown in Fig.~\ref{fig:nw_gap}.
The difference between 6 fm$^{-1}$ and 10 fm$^{-1}$ can be partially
attributed to the finite-point approximation to the derivatives,
which underestimates the kinetic energy by 30\% at $k=\frac{\pi}{a}$,
as shown in Fig.~20 of Ref.~\cite{Taj01a}. 

Incidentally, the requirement of orthogonality among the canonical orbitals
does not seem to change the conclusion drawn with this model.
The first reason is that point collapses in light nuclei
are not affected by orthogonality
if the number of canonical orbitals $N_{\rm w}$ is relatively small.
This is because of the symmetry (e.g., $D_{2h}$) of the solution:
If a high-lying orbital is the only one in an irreducible representation
of the symmetry (e.g., in $P_x$=$-$ sector when $x$-axis is the shortest
axis), it is automatically orthogonal to the other orbitals.
In this case, it shrinks not to a point but to
many points (e.g., two points at $x=\pm R$, $y=z=0$, 
$R$ $\sim$ nuclear radius, with opposite-sign amplitudes).
Indeed, we have experienced that point collapses occur 
only when $N_{\rm w}$ is small for relatively small values of $k_{\rm c}$.
The second reason is an empirical fact that,
with large values of $k_{\rm c}$, many orbitals collapse
one after another or simultaneously even for large $N_{\rm w}$:
The process of point collapses under orthogonality condition
may not be so complicated as it sounds in such cases.

%----- effects of the tilde{rho}_c term ------------------------------

The second type of the interaction we examine is the pairing density 
dependent term of Eq.~(\ref{eq:pair_int}).
With $\tilde{\rho}_{\rm c} < \infty$, but with $k_{\rm c} = \infty$,
$\Delta E_{\rm tot}$ has additional terms to Eq.~(\ref{eq:DEtot}) as,
\begin{equation} \label{eq:DEtotrp}
\Delta E_{\rm tot}'' = \Delta E_{\rm tot}
                     + C_{13} uv 
                     + C_{14} \frac{u^2 v^2}{w^3}
                     + C_{15} \frac{u^3 v^3}{w^6}
                     + C_{16} \frac{u^4 v^4}{w^9} ,
\end{equation}
where
\begin{eqnarray}
C_{13} & = & -\frac{2 v_{\rm p}}{\tilde{\rho}_{\rm c}^2}
             \tilde{\rho}_0^3 , \nonumber \\
C_{14} & = & -\frac{2 v_{\rm p}}{\tilde{\rho}_{\rm c}^2}
       \frac{6}{\left(\sqrt{\pi}\right)^3} \tilde{\rho}_0^2 , \nonumber \\
C_{15} & = & -\frac{2 v_{\rm p}}{\tilde{\rho}_{\rm c}^2}
       \frac{128}{\left(\sqrt{3}\pi\right)^3} \tilde{\rho}_0 , \nonumber \\
C_{16} & = & -\frac{2 v_{\rm p}}{\tilde{\rho}_{\rm c}^2}
             \frac{64}{\sqrt{2\pi^9}} . \nonumber
\end{eqnarray}
All the added terms are positive. 
Assuming $v = \alphaip w^{\betaip}$,
terms with coefficients $C_{15}$ and $C_{16}$ have negative
powers of $w$ at $\betaip = \frac{3}{2}$. 
Consequently point collapses can be avoided
by using finite values of $\tilde{\rho}_{\rm c}$.

%-------------------------- F I G U R E -----------------------------------
%\begin{figure}[htb]%G:box
%\begin{center}\framebox{Figure \ref{fig:shrinkrp}}\end{center}%G:box
\begin{figure}%G:psf
\includegraphics[angle=-90, width=8.6 cm]{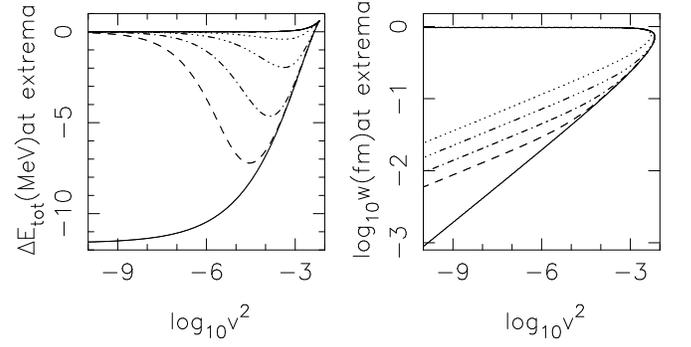}%G:psf
\caption{
Same as in Fig.~\figshrinkkc , but for the effects of $\tilde{\rho}_{\rm c}$.
The dot, triple-dot-dash, dot-dash, dash, and solid lines are obtained with
$\tilde{\rho}_{\rm c} = 1.5, 6, 25, 100$, and $\infty$ (fm$^{-3}$), 
respectively.
The other parameters of the model are the same as those used for
Fig.~\figshrink.
}
\label{fig:shrinkrp}
\end{figure}
%------------------- E N D   O F   F I G U R E ----------------------------

Figure~\ref{fig:shrinkrp} shows extrema of
$\Delta E_{\rm tot}$ versus $v^2$.
The solid curve corresponds to $\tilde{\rho}_{\rm c} = \infty$
and is the same as that in Fig.~\ref{fig:shrinkkc}.
The other types of curves are for finite values of $\tilde{\rho}_{\rm c}$.
One can see that $\Delta E_{\rm tot}$ at the local minimum
is already very small with $\tilde{\rho}_{\rm c} = 1.5 $ fm$^{-3}$ (dot curve),
which is much larger than the typical values of $\tilde{\rho}$
and thus leads to only physically acceptable small amount of modification
of the pairing interaction.
This figure is consistent with the results of actual HFB calculations:
From Fig.~4 of Ref.~\cite{Taj00a}, 
one can see that point collapses can be avoided with
$\tilde{\rho}_{\rm c} < 1.0$ fm$^{-3}$.

Mathematically speaking, however, the local minimum continues
to exist for arbitrary $v$,
which is a different behavior from that with finite $k_{\rm c}$.
In the limit of $v \rightarrow 0$, this minimum behaves as
$w \rightarrow 0$ and $\Delta E_{\rm tot}'' \rightarrow -0$.
Because the depth converges to zero from below, 
the point-collapse limit is unstable and is not very meaningful.
So, let us give only briefly the result of an analytical consideration.
Assuming $v = \alphaip w^{\betaip}$,
one can show that
the local minimum continues to exist in the limit $w \rightarrow 0$
with $\betaip = 3$ and $\alphaip$ determined by a cubic equation
$C_1 + C_{13} +2 C_4 \alphaip + 4 C_{16} \alphaip^3 = 0$
if $C_1 + C_{13} <0$
(i.e., $\tilde{\rho}_{\rm c} > \sqrt{2} \tilde{\rho}_0$
for $\rho_0 \ll \rho_{\rm c}$).
At the local minimum, $\Delta E_{\rm tot}'' \propto v$ and $w \propto v^{1/3}$.
One can see this scaling behavior with $\betaip=3$
on the right-hand side of Fig.~\ref{fig:shrinkrp}.

In the present paper we have used the momentum dependent interaction
because it has a physical origin, i.e., the finite-rangeness.
However, the pairing density dependent interaction may also be
useful under different circumstances because the momentum dependent
interaction leads to a divergence of the pairing energy in
unrestricted Hilbert space \cite{TOT94}: The assumption of $uv \ge 0$
made in this paper is not adequate for orbitals with $k > k_{\rm c}$.

% summary

To summarize, assuming a wave function of Gaussian form, we have
clarified the mechanism of the point-collapse phenomenon occurring in
the canonical-basis HFB method due to delta interactions. We have also
demonstrated the successes of momentum or pairing density dependent
interactions in avoiding the collapse.

%==============================================================================

%----------------------------------------------------------------------------

\end{document}